\documentclass[%
reprint,
superscriptaddress,
secnumarabic,
 amsmath,amssymb,
 aps,
 prx,
]{revtex4-2}

\usepackage[euler]{textgreek}
\usepackage[utf8]{inputenc}
\usepackage[margin=0.75in]{geometry}
\usepackage{graphicx}
\usepackage{bbold}
\usepackage{xcolor}
\usepackage{cleveref}

\everymath{\displaystyle}

\newcommand{\singletS}{${}^1$S$_0\,$}
\newcommand{\tripletS}{${}^3$S$_1\,$}
\newcommand{\tripletPone}{$^3$P$_1\,$}
\newcommand{\tripletPzero}{$^3$P$_0\,$}

\newcommand{\ybOne}{$^{171}$Yb}
\newcommand{\ybFour}{$^{174}$Yb}
\usepackage{physics}
\usepackage{enumitem}
\begin{document}
\title{Universal gate operations on nuclear spin qubits in an optical tweezer array of $^{171}$Yb atoms}
\author{Shuo Ma}
\thanks{These authors contributed equally to this work.}
\affiliation{Princeton University, Department of Electrical and Computer Engineering, Princeton, New Jersey 08544}
\affiliation{Princeton University, Department of Physics, Princeton, New Jersey 08544}
\author{Alex P. Burgers}
\thanks{These authors contributed equally to this work.}
\affiliation{Princeton University, Department of Electrical and Computer Engineering, Princeton, New Jersey 08544}
\author{Genyue Liu}
\affiliation{Princeton University, Department of Electrical and Computer Engineering, Princeton, New Jersey 08544}
\author{Jack Wilson}
\affiliation{Princeton University, Department of Electrical and Computer Engineering, Princeton, New Jersey 08544}
\author{Bichen Zhang}
\affiliation{Princeton University, Department of Electrical and Computer Engineering, Princeton, New Jersey 08544}
\author{Jeff D. Thompson}
\email[ ]{jdthompson@princeton.edu}
\affiliation{Princeton University, Department of Electrical and Computer Engineering, Princeton, New Jersey 08544}

\begin{abstract}
Neutral atom arrays are a rapidly developing platform for quantum science. In this work, we demonstrate a universal set of quantum gate operations on a new type of neutral atom qubit: a nuclear spin in an alkaline earth-like atom (AEA), $^{171}$Yb. We present techniques for cooling, trapping and imaging using a newly discovered magic trapping wavelength at $\lambda = 486.78$ nm. We implement qubit initialization, readout, and single-qubit gates, and observe long qubit lifetimes, $T_1 \approx 20$ s and $T^*_2 = 1.24(5)$ s, and a single-qubit operation fidelity $\mathcal{F}_{1Q} = 0.99959(6)$. We also demonstrate two-qubit entangling gates using the Rydberg blockade, as well as coherent control of these gate operations using light shifts on the Yb$^+$ ion core transition at 369 nm. These results are a significant step towards highly coherent quantum gates in AEA tweezer arrays.
\end{abstract}

\maketitle
\section{Introduction}
Neutral atom arrays with Rydberg-mediated entanglement are a versatile and rapidly developing tool for quantum computing and simulation \cite{saffman2016,browaeys2020}. In recent years, this approach has been used to study quantum phase transitions \cite{ebadi2020, scholl2020} and probe novel phases of matter \cite{semeghini2021} in large, two-dimensional systems. At the same time, the performance of quantum gate operations is also improving rapidly \cite{graham2019,Levine2019}, allowing the generation of many-qubit entanglement \cite{omran2019} and new approaches to quantum state tomography and benchmarking \cite{choi2021}.

An emerging frontier within neutral atom arrays is the use of alkaline earth-like atoms (AEAs) such as Sr \cite{Cooper2018,Norcia2018,barnes2021} and Yb \cite{Saskin2019}. The rich internal structure of these atoms affords numerous unique capabilities, including ground-state Doppler cooling \cite{Cooper2018,Norcia2018}, trapping of Rydberg states \cite{Wilson2019}, high-fidelity single-photon Rydberg excitation \cite{madjarov2020} and efficient local control of gate operations using light shifts on the ion core \cite{burgers2021,pham2021}. Furthermore, the presence of optical clock transitions opens new applications to frequency metrology \cite{Norcia2019,madjarov2019}, particularly using entangled states \cite{schine2021}.

Another unique capability of fermionic AEA isotopes (with odd masses) is encoding qubits in the nuclear spin of the $J=0$ electronic ground state. Because of the absence of hyperfine coupling, this qubit should be highly immune to many sources of technical noise, including differential light shifts \cite{kuhr2005} and Raman scattering \cite{dorscher2018}, while still allowing high-fidelity operations using hyperfine coupling in electronic excited states. Nuclear spin qubits have been experimentally demonstrated in $^{87}$Sr tweezer arrays ($I=9/2$) \cite{barnes2021}, and in single $^{171}$Yb atoms ($I=1/2$) trapped in an optical cavity \cite{noguchi2011}. However, a universal set of quantum logic operations, including two-qubit gates through a Rydberg state, has not been demonstrated for nuclear spin qubits in AEAs.

In this work, we present the first such demonstration, using qubits encoded in the nuclear spin of \ybOne. We present high-fidelity imaging of \ybOne\ atoms in a 30-site optical tweezer array using a newly discovered magic wavelength, as well as techniques for initialization, readout and one- and two-qubit gate operations. Crucially, the qubit coherence time, $T^*_2 = 1.24(5)$ s, exceeds most recent measurements of hyperfine state coherence times in alkali atom tweezer arrays by 2-3 orders of magnitude \cite{beugnon2007,xia2015,Levine2019,graham2019}, because of the dramatically reduced differential light shift from the trap (a notable exception is \cite{guo2020}, where $T^*_2 = 0.94(3)$ s for Rb is realized using a circularly-polarized, magic-intensity trap). The depolarization time, $T_1 = 20(2)$ s (corrected for atom loss), is also significantly improved compared to alkali atoms. These properties enable single-qubit gates with an average fidelity $\mathcal{F}_{1Q} = 0.99959(6)$. We also implement two-qubit gates using the Rydberg blockade, which acts on the nuclear spin through the hyperfine coupling in the Rydberg manifold, and generate Bell pairs with fidelity $\mathcal{F}_B = 0.83(2)$. Lastly, we demonstrate coherent control over the Rydberg gate using light shifts on an optical transition in the Yb$^+$ ion core. These results provide a foundation for large-scale, high-fidelity quantum logic and simulation using nuclear spin qubits in AEA tweezer arrays, as well as tweezer clocks using fermionic AEA isotopes \cite{lemke2009}.

\begin{figure}[ht]
\centering
\includegraphics[width=1\linewidth]{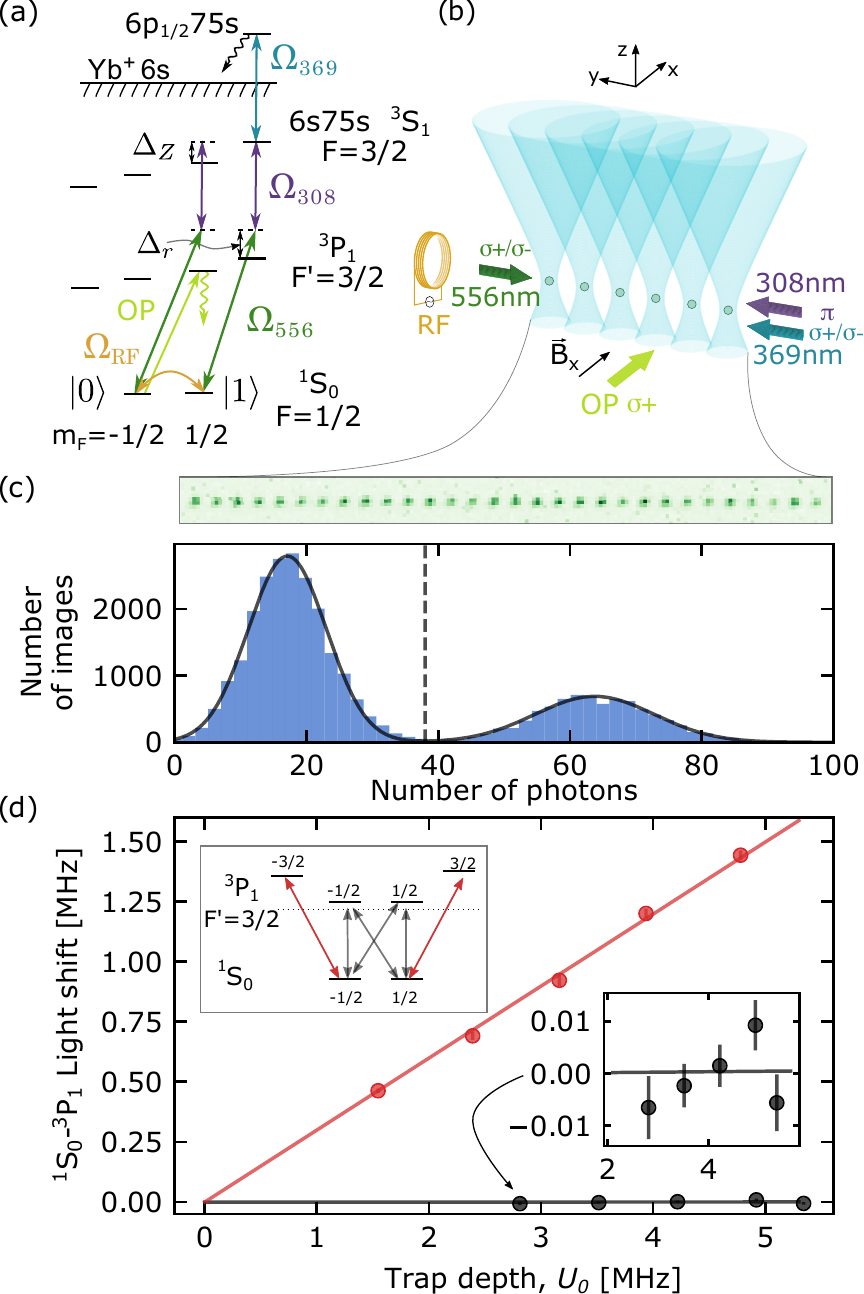}
\vspace{-0.5cm}
\caption{\label{fig:overview}  (a) $^{171}$Yb level diagram. A qubit is encoded in the nuclear spin sublevels of the \singletS ground state, and initialized into $\ket{1}$ via optical pumping (OP). Spin rotations are performed using an RF field. Two-qubit gates are performed using the $6s75s$ $^3$S$_1$ $\ket{F,m_F} = \ket{3/2,3/2}$ level, accessed via two-photon excitation through $^3$P$_1$. State readout is performed by blowing out atoms in $\ket{1}$ by autoionization from the Rydberg state. (b) Orientation of the control fields and tweezers. (c) Histogram of photons detected in a single tweezer region during a 25 ms image, and a fluorescence image of a 30-site array (averaged over many exposures). (d) Differential light shift measurement on the \ybOne\ \singletS-\tripletPone\ transition in tweezers with $\lambda = 486.78$ nm. The $F'=3/2$ manifold is split by the tensor light shift, giving a differential shift of $0.3\,U_0$ for $|m_{F'}|=3/2$, and $0.0001(6)\,U_0$ for $|m_{F'}|=1/2$.}
\vspace{0mm}
\end{figure}

\section{Imaging $^{171}$Yb atoms in tweezers}
High-fidelity imaging of atoms in optical tweezers requires bright fluorescence and efficient cooling. The narrow \singletS$\rightarrow$\tripletPone transition ($\Gamma=2\pi\times182$ kHz, $\lambda=556$ nm) provides a good balance of high scattering rate and low Doppler temperature, and was recently used to image \ybFour\ atoms in an array of optical tweezers. Imaging with a narrow linewidth benefits from a state-insensitive trap at a magic wavelength, which is achieved at $\lambda=532$ nm for the $\pi$-polarized imaging transition in \ybFour\ \cite{yamamoto2016, Saskin2019}. However, this wavelength is no longer magic in \ybOne\ because of the interplay of the tensor light shift and the hyperfine structure. As a result, we observed poor imaging performance in \ybOne\ with this trapping wavelength.

To overcome this challenge, we searched for a new magic wavelength for this transition in \ybOne. Using literature dipole matrix elements \cite{porsev1999} and measurements in \ybFour\ tweezer arrays at various wavelengths, we predicted that a magic wavelength for \ybOne\ transition should occur near 486 nm (see Appendix \ref{app:pol}). We verified this prediction using a direct measurement with \ybOne\ atoms. In an optical tweezer array at $\lambda = 486.78$ nm (in vacuum), we find a light shift of $0.0001(6)\,U_0$ for the $F'=3/2$ $|m_{F'}|=1/2$ transitions, compared to $0.3\,U_0$ for the $|m_{F'}|=3/2$ transitions, where $U_0$ is the ground state trap depth. From this measurement at a single wavelength, we estimate that the exact magic wavelength is in the range $\lambda = 486.78 \pm 0.1$ nm.

At this wavelength, we obtain imaging and cooling performance comparable to \ybFour. In a 30-site array using a power of $5$ mW per tweezer, we perform repeated imaging of single atoms with a combined fidelity and survival probability between images of 98.8(3)\% (Fig. \ref{fig:overview}c), using an exposure time of $25$ ms. The atomic temperature during imaging is $T=7.4(7)$ $\mu$K.

We have also observed vanishing light shifts for the $^3$P$_1$ $|m_{F'}|=1/2$ transition in an optical tweezer at the clock magic wavelength transition of 759.35 nm \cite{barber2008}, which is promising for future experiments on tweezer clocks \cite{Norcia2019,madjarov2019} using $I=1/2$ hyperfine isotopes \cite{lemke2009}.

\section{Single qubit operations}\label{sec:sq}

We encode qubits in the sublevels of the $F=1/2$ \singletS\ level, and label the states $\ket{m_F=-1/2} \equiv \ket{0}$ and $\ket{m_F=1/2}\equiv\ket{1}$ (Fig. 1a). Since there is no hyperfine coupling in \singletS, initialization and readout rely on the hyperfine and Zeeman shifts of the excited states (Fig. \ref{fig:overview}a). State initialization is performed using optical pumping (OP) with $\sigma^+$-polarized light, on the $\ket{0}\rightarrow \ket{^3P_1, F'=3/2,m_{F'}=1/2}$ transition. A magnetic field ($B_x=4.11$ G) shifts the $m_{F'}=3/2$ transition by $5.7$ MHz $=31 \Gamma$, ensuring that $\ket{1}$ is a dark state.

Spin readout is performed by selectively blowing out atoms in $\ket{1}$, via autoionization from the Rydberg state. We use the $^3$S$_1$ $6s75s$ Rydberg state $\ket{F'=3/2,m_{F'}=3/2}$, which has not been previously observed, to the best of our knowledge (Appendix \ref{app:blockade}). We excite this state in a two-photon process through $^3$P$_1$, with intermediate state detuning $\Delta_r=2\pi \times 20$ MHz. The two-photon Rabi frequency is $\Omega_r=2\pi\times0.31$ MHz. Excitation of $\ket{0}$ is prevented by a Zeeman splitting in the Rydberg manifold, $\Delta_Z=2\pi\times7.8$ MHz $\gg \Omega_r$. Following a $\pi$ pulse on this transition, any Rydberg atoms are rapidly autoionized by exciting the $6s\rightarrow6p_{1/2}$ (369 nm) Yb$^+$ ion core transition \cite{Lochead2013,madjarov2020,burgers2021}, and atoms remaining in $\ket{0}$ are detected via fluorescence imaging.

We characterize the detection protocol by performing $N_r$ blowout cycles before imaging (Fig. 2a). Because of imperfect Rydberg $\pi$ pulses, some atoms in $\ket{1}$ remain after a single round, but this population is suppressed exponentially with additional rounds. With $N_r = 3$, we detect a surviving atom with probability of $94.5(3)\%$ ($0.2(2)\%$) after preparation in $\ket{0}$ ($\ket{1}$), corresponding to an average initialization and readout fidelity of 97.2(4)\%. We attribute this mainly to imaging errors (0.6(2)\%) and Raman scattering during the Rydberg excitation (1.2(2)\%).

\begin{figure}[tp]
\centering
\includegraphics[width=1\linewidth]{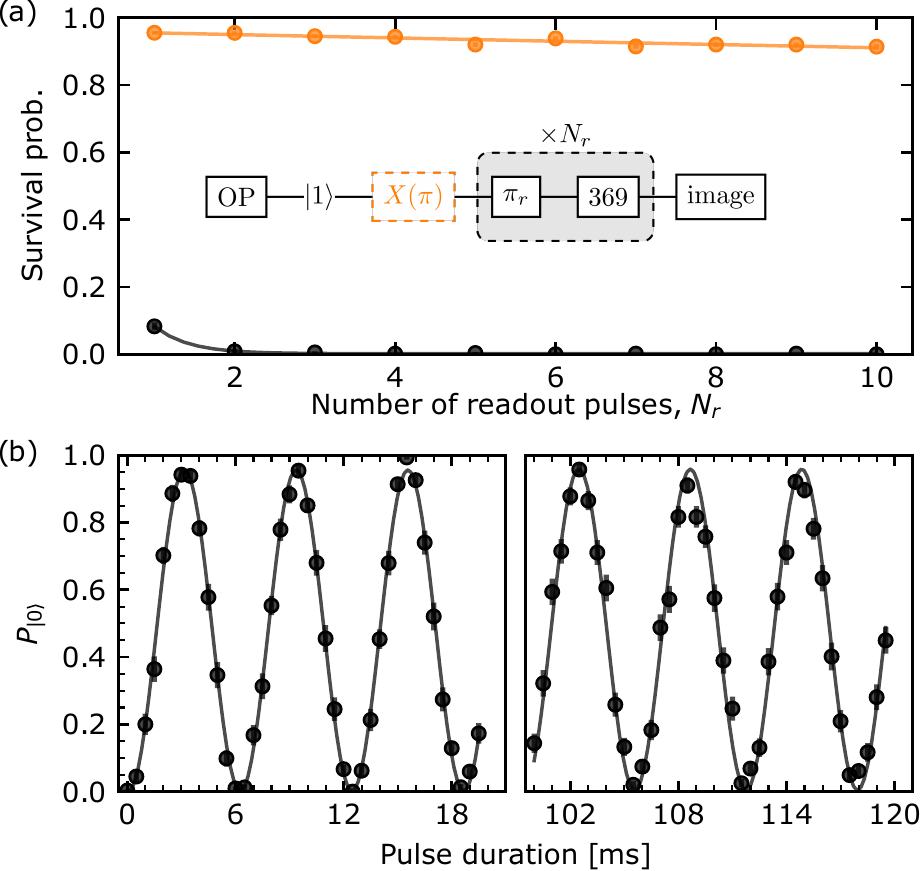}
\vspace{-5mm}
\caption{\label{fig:readout}
(a) Atom survival probability after $N_r$ repetitions of the blowout sequence through the Rydberg state, after preparing an atom in $\ket{1}$ (black) and $\ket{0}$ (orange). (b) Rabi oscillations of the nuclear spin qubit, with $\Omega_{RF} = 2\pi\times161.7(5)$ Hz.
}
\vspace{0mm}
\end{figure}

Next, we demonstrate coherent manipulation of the nuclear spin qubit. The magnetic field splits the nuclear spin hyperfine levels by $750$ Hz/G, resulting in a Larmor frequency $\omega_L \approx 2\pi \times 3.09$ kHz. This enables qubit rotations using an RF coil. We observe long-lived nuclear spin Rabi oscillations with Rabi frequency $\Omega_{RF}=2\pi\times161.7(5)$ Hz (Fig. \ref{fig:readout}b). The small value of $\Omega_{RF}$ is chosen to avoid counter-rotating terms when $\Omega_{RF} \approx \omega_L$. Faster rotations are readily accessible with larger static fields, or by implementing a second coil to generate $\sigma^+$-polarized RF.

To quantify the qubit coherence, we perform a Ramsey measurement. Fitting the data from each site individually yields an average $T_{2}^*=1.24(5)$ s. Averaging the populations across the array before fitting (Fig. 3a) results in a slightly lower value $\bar{T}_2^*=1.00(4)$ s, because of a magnetic field gradient giving a shift of $\approx2$ mHz/$\mu$m in the array direction.

To mitigate slowly-varying noise and inhomogeneity, we perform a spin-echo measurement. At each hold time, we vary the phase of the second $\pi/2$ pulse and extract the coherence from the visibility of the resulting oscillation (Fig. 3b), removing the effect of magnetic field gradients and shot-to-shot fluctuations. The raw visibility decays to $1/e$ after 3.9(3) s. Normalizing to the mean survival probability to remove the effect of atom loss gives an estimate of the intrinsic $1/e$ decoherence time of $T_2=5(1)$ s. We attribute this residual decoherence to magnetic field drifts during each experiment, arising from thermal effects in our current source (which are also visible as a chirp in the data in Fig. 3a).

We also measure the depolarization time, $T_1$, after initializing in each spin state (Fig. 3c). Atom loss occurs on a timescale $T_a = 9$ s. By normalizing the spin state population to the fraction of atoms remaining, we extract the intrinsic spin flip rates starting from each state: $T_1^{\ket{0}}=13(2)$ s and $T_1^{\ket{1}}=27(2)$ s. In the absence of hyperfine coupling in  $^1$S$_0$, the predicted Raman scattering rate is extremely small, of order $10^{-13}$ s$^{-1}$ \cite{dorscher2018}, and we attribute the observed spin flip rate to leaked resonant light from a beam path without a mechanical shutter.

\begin{figure}[tp]
\centering
\includegraphics[width=1\linewidth]{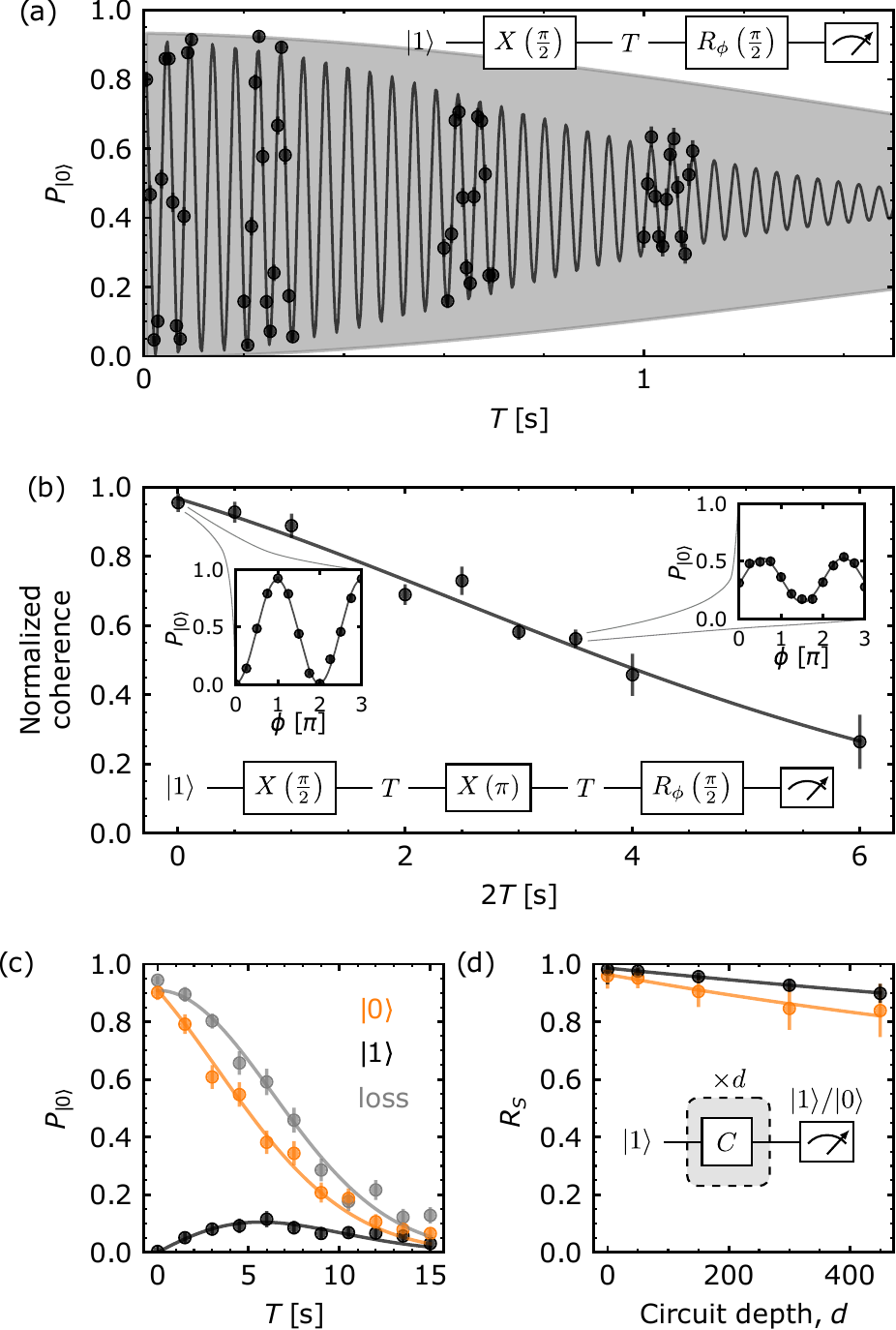}
\vspace{-5mm}
\caption{\label{fig:coherence}Coherence measurements of the $^{171}$Yb nuclear spin qubit. (a) Array-averaged Ramsey fringes with fitted $1/e$ decay time of $\bar{T}_2^*=1.00(4)$ s (fitting sites individually yields an average $T_2^* = 1.24(5)$ s). The gray shaded envelope indicates the contribution from a magnetic field gradient across the array. (b) Spin echo measurement. The phase of the final $\pi/2$ pulse is scanned for each wait time $T$, and the coherence is extracted from the visibility (insets) and normalized to the mean survival. The normalized coherence decays to $1/e$ after a time $T_2=5(1)$ s. (c) Spin $T_1$ measurement. The orange (black) points show the probability to detect an atom in $\ket{0}$ following initialization in $\ket{0}$ ($\ket{1}$). The gray points show survival probability without the spin readout sequence (\emph{i.e.}, atom loss), which decays to $1/e$ after $T_a = 9.0(3)$ s. Using survival-normalized populations, we infer an intrinsic $1/e$ spin flip time for each state: $T_1^{\ket{0}}=13(2)$ s and $T_1^{\ket{1}}=27(2)$ s. (d) Randomized benchmarking of single qubit rotations. Orange (black) points show the probability to obtain the correct result for sequences ending in $\ket{0}$ ($\ket{1}$). From the decay rates, we obtain an average gate fidelity of 99.959(6)\%.}
\vspace{0mm}
\end{figure}

Lastly, we study the single-qubit gate fidelity with randomized benchmarking (RB). We use pyGSTi \cite{Nielsen2020} to generate random sequences of Clifford group operations of varying depth (40 sequences per depth). The ideal output of the RB circuit is chosen to be $\ket{0}$ or $\ket{1}$ with equal probability. The probability to get the correct output, $R_S$, is measured and fitted (Fig. \ref{fig:coherence}d) to extract the average operation fidelity according to:
\begin{equation}
   R_S=\frac{1}{2}+\frac{1}{2}(1-2\epsilon_{\rm{spam}})(1-2\epsilon_{\rm gate})^d,
\end{equation}
\noindent where $\epsilon_{\rm{spam}}$ is the state preparation and measurement error, $\epsilon_{\rm gate}$ is the average gate error, and $d$ is the circuit depth. We find the average gate fidelity $\mathcal{F}_{1Q}=1-\epsilon_{\rm gate}$ to be 99.975(1)\% and 99.927(3)\% using RB circuits with an ideal output state in $\ket{1}$ and $\ket{0}$, respectively, for an average fidelity of 99.959(6)\%. The difference in fidelity between the two output states is consistent with atom loss, given the sequence duration of $d \times 4$ ms. The average $\epsilon_{\rm{spam}}=0.025(6)$.

\section{Two-qubit entangling gate}\label{sec:2q}

To complete the universal set of operations, we implement a two-qubit gate based on the Rydberg blockade \cite{lukin2001}. A number of blockade-based gate protocols have been demonstrated \cite{isenhower2010, jau2016, wilk2010, Levine2019, graham2019}, based on selective coupling of one of the qubit levels to a Rydberg state. In our work, we realize selective coupling of $\ket{1}$ to the Rydberg state by applying a magnetic field that produces a large Zeeman shift $\Delta_z \gg \Omega_r$ in the Rydberg state (Section \ref{sec:sq}, Fig. 1a). In this arrangement, the nuclear spin is manipulated using the hyperfine coupling in the $F'=3/2$ Rydberg state. For the two-qubit gate protocol we use a larger Rabi frequency $\Omega_r=2\pi\times$0.763 MHz ($\Delta_Z/\Omega_r = 10.6$).

\begin{figure}[h!]
\centering
\includegraphics[width=1\linewidth]{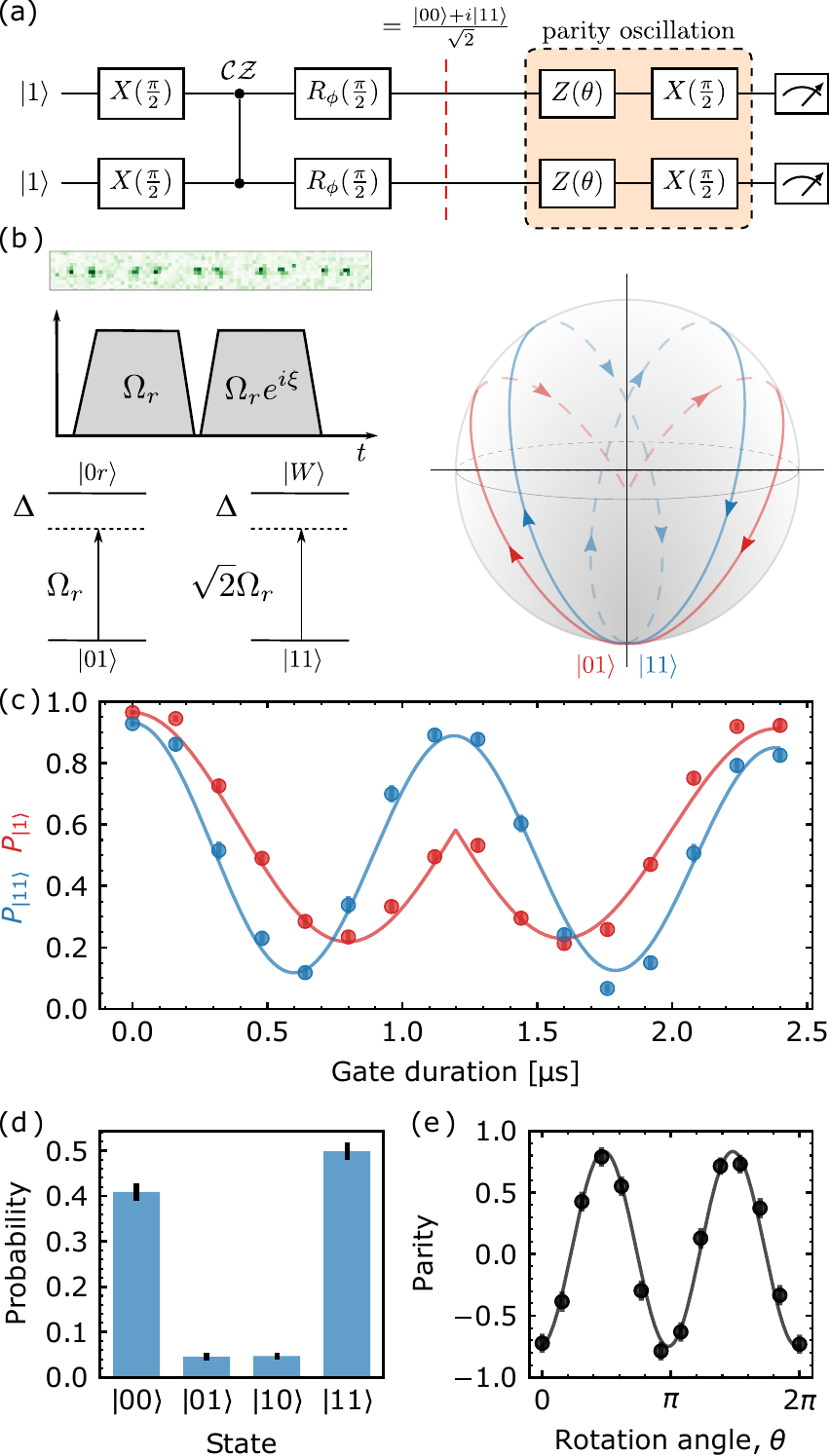}
\vspace{-5mm}
\caption{\label{fig:gate}Two-qubit gate. (a) Circuit diagram of the two-qubit gate characterization experiment. (b) Single shot image of 5 dimers used for the two-qubit gate protocol (the atom spacing is increased for imaging). The gate consists of two Rydberg pulses, with a phase shift on the second pulse. The level diagrams and Bloch sphere trajectories indicate the dynamics for the computational states $\ket{01}$ (red) and $\ket{11}$ (blue). (c) Evolution of the states $\ket{1}$ (blue, used as a proxy for $\ket{01}$) and $\ket{11}$ (red) state during the gate. The solid lines are simulations using the master equation with a phenomenological dephasing contribution. (d) Measured state populations at the circuit location indicated by the red dashed line. (e) Parity oscillations, with contrast $\mathcal{C} = 0.79(3)$.}
\vspace{0mm}
\end{figure}

We implement the controlled-Z (CZ) gate protocol of Ref. \cite{Levine2019}, because it acts symmetrically on the atoms and does not require local addressing of the Rydberg excitation light. The gate consists of two pulses of length $\tau$, detuning $\Delta$, and Rabi frequency $\Omega_r$, but with a phase slip $\xi$ on the second pulse (Fig. 4b). The behavior can be understood by considering the evolution of the computational basis states. The state $\ket{00}$ does not undergo any dynamics. The state $\ket{01}$ ($\ket{10}$) undergoes detuned Rabi oscillations to $\ket{0r}$ ($\ket{r0}$). The state $\ket{11}$ undergoes detuned Rabi oscillations to $\ket{W}=(\ket{1r}+\ket{r1})/\sqrt{2}$, with a Rabi frequency of $\sqrt{2}\Omega_r$, since the Rydberg blockade suppresses excitation to $\ket{rr}$. The difference in Rabi frequencies allows the pulse detuning and duration to be chosen to return both $\ket{01}$ and $\ket{11}$ to themselves at the end of the gate (Fig. 4c), but with distinct phase accumulations:

\begin{align*}
    \ket{00} & \rightarrow \ket{00}, \\
    \ket{01} & \rightarrow \ket{01} e^{i\phi_{01}}, \\
    \ket{10} & \rightarrow \ket{10} e^{i\phi_{10}}, \\
    \ket{11} & \rightarrow \ket{11} e^{i\phi_{11}}.
\end{align*}

\noindent For $\Delta/\Omega_r \approx 0.377$ and $\xi \approx 3.902$ and $\tau \Omega_r \approx 4.293$, $\phi_{11} = 2\phi_{01} + \pi$, realizing a CZ gate \cite{Levine2019}. 

We experimentally implement the CZ gate in parallel on five pairs of atoms in a dimerized array, generated deterministically using rearrangement. The atomic separation within a dimer is 3 $\mu$m, considerably less than the measured blockade radius of 14(1.4) $\mu$m (see Appendix \ref{app:blockade}), while the spacing between dimers is 30 $\mu$m. The state dynamics during the gate are depicted schematically in Fig. \ref{fig:gate}b, and measured in Fig. \ref{fig:gate}c.

To determine the fidelity of the complete gate, we use the circuit in Fig. 4a. We prepare the two-atom state $\left[(\ket{0} - i\ket{1})/\sqrt{2}\right]^{\otimes 2}$, apply the CZ gate, then drive a $\pi/2$ pulse to produce the Bell state $\ket{\phi} = (\ket{00} + i\ket{11})/\sqrt{2}$. The rotation axis of the second pulse is at an angle $\phi = -6.6$ rad with respect to the initial pulse, to compensate the single-qubit phase $\phi_{01}$ and a light shift from 556 nm beam. The fidelity is determined from the state populations (Fig. \ref{fig:gate}d) and the contrast $\mathcal{C}$ of parity oscillations (Fig. \ref{fig:gate}e) following an additional single-qubit rotation, according to \cite{sackett2000}:
\begin{equation}\label{eqn:fidelity}
  \mathcal{F}_{B} = \frac{1}{2}(p_{00}+p_{11}+\mathcal{C}),  
\end{equation}

\noindent where $p_{00}$ and $p_{11}$ are the populations in $\ket{00}$ and $\ket{11}$.

We obtain a raw Bell state fidelity of $\mathcal{F}_{B}=85(2)$\%. This value is affected by errors in the state preparation and measurement (SPAM), as well as leakage to the Rydberg state. Correcting for these effects with independent measurements, we determine a conservative lower bound on the Bell state fidelity, $\mathcal{F}_B^c \geq 83(2)\%$ (Appendix \ref{app:fidelity}). This is limited by several factors, including spontaneous emission from the intermediate and Rydberg states ($\approx 3\%$), as well as laser noise and Doppler decoherence, which are more problematic for the relatively small value of $\Omega_r$ used here \cite{deleseleuc2018}.

\section{Controlling the two-qubit gate}\label{sec:ls}

Lastly, we demonstrate coherent control of the two-qubit gate operation using a light shift on $\ket{r}$ arising from the the Yb$^+$ ion core transition ($\lambda = 369$ nm), which is driven by a (global) control field \cite{burgers2021}. By shifting the $\ket{1}\rightarrow\ket{r}$ transition out of resonance with the Rydberg excitation beam during the CZ gate (Fig. 5a), the action of the gate is suppressed (Fig. 5b). In this experiment, the control field intensity is approximately 700 W/cm$^2$, and it is detuned by $\Delta_c=-2\pi\times 5$ GHz from the $6s75s\rightarrow6p_{1/2}75s$ (autoionization) transition. We measure a light shift of -18.3(1) MHz on $\ket{r}$.

We quantify the coherence of this approach to gate control in two ways. First, we implement the circuit in Fig. 5a, where the suppressed CZ gate should result in the identity operation (plus a single-qubit phase, compensated by the second single-qubit pulse). The ideal circuit output is then $\ket{00}$, and we can quantify the state fidelity by measurement in the population basis (Fig. 5c). We obtain a raw fidelity of $\mathcal{F}_I = 84(2)\%$, or $\mathcal{F}^c_I = 93(3)\%$ after SPAM correction (Appendix \ref{app:fidelity}). We attribute the infidelity to Raman scattering ($4(2)\%$) from the 556 nm gate beam, and autoionization from a small Rydberg population that is present despite the control beam ($2(1)\%$). The former can be suppressed using single-photon excitation (\emph{i.e.}, starting in $^3$P$_0$), while the latter can be suppressed by higher control beam power or detuning \cite{burgers2021}.

Second, we directly measure the qubit decoherence and depolarization induced by the control beam. We perform Ramsey, spin echo and $T_1$ measurements with continuous illumination from the control beam, and find zero effect in all cases (\emph{e.g.}, Fig. 5d). We also measure the differential light shift on qubit levels resulting from the control beam, and obtain a value of $40(14)$ mHz, approximately $10^{-9}$ times smaller than the light shift of the Rydberg state under the same conditions. This indicates that the control beam has minimal perturbation on the ground state qubits.

\begin{figure}[]
\centering
\includegraphics[width=1\linewidth]{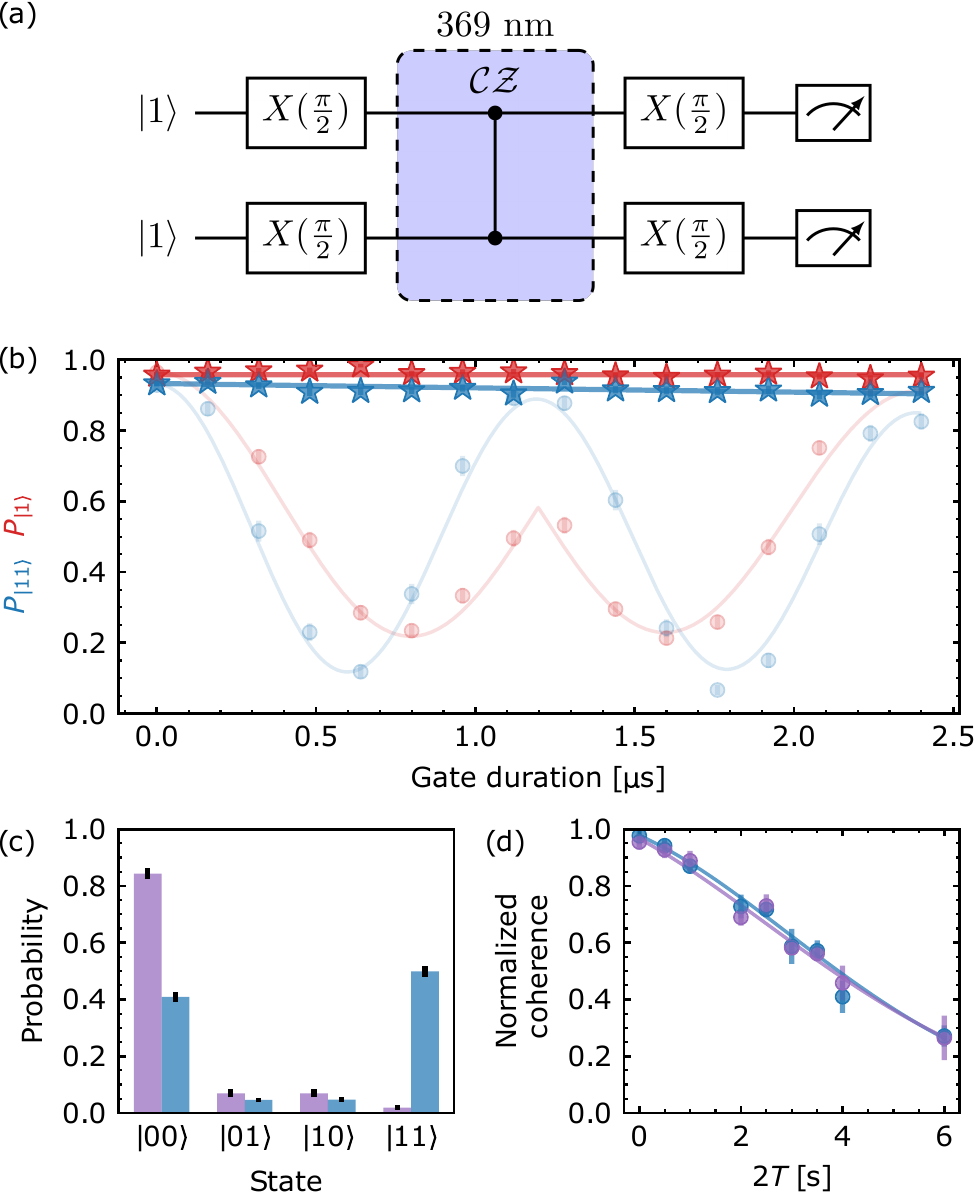}
\vspace{-5mm}
\caption{\label{fig:autionization} Coherent control of a Rydberg gate using an ion core light shift. (a) The two-qubit gate is suppressed by applying light near resonance with a Yb$^+$ ion core transition (369 nm) during the gate. (b) Evolution of the states $\ket{1}$ (red stars) and $\ket{11}$ (blue stars) during a suppressed gate. The dynamics without the control light are shown for comparison (reproduced from Fig. 4c). (c) Measured state populations with (purple) and without (blue) the control light. (d) Coherence of the nuclear spin in \singletS\ under continuous illumination with the control light (purple). The measured $T_2 = 5.1(7)$ s is consistent with Fig. 3b (reproduced here, in blue).}
\vspace{0mm}
\end{figure}

\section{Discussion}

The long coherence times observed in this work are important not only for improved gate fidelity, but also for taking advantage of other aspects of the optical tweezer architecture, such as preserving qubit coherence while moving atoms \cite{beugnon2007,yang2016}. In alkali atoms in optical tweezers, $T_2^*$ is often limited to $1-2$ ms by fractional differences in the trap potentials for the qubit states of order $\eta \approx 10^{-3}$ \cite{beugnon2007,xia2015,graham2019,levine2021} (a notable exception is \cite{guo2020}, using hyperpolarizability to realize magic intensity traps with $\eta=0$, and $T_2^* = 0.94(3)$ s). For nuclear spin qubits, we expect $\eta=0$ for linearly polarized light, and experimentally measure $\eta=9.8\times 10^{-8}$. This would limit $T_2^*$ to 15 s, implying another origin for the observed decoherence, such as magnetic field instabilities. Alkali atoms in tweezers also experience Raman scattering that leads to depolarization on a timescale $T_1 \approx 0.5-1$ s \cite{xia2015, graham2019,levine2021}. In \ybOne\, this should be very strongly suppressed by the absence of hyperfine coupling in $^1$S$_0$, and we predict Raman transition rate of $10^{-13}$ s$^{-1}$ for our parameters \cite{dorscher2018}. We believe that the observed value, $T_1 \approx 20$ s, is limited by resonant light leakage. We note that longer coherence times and $T_1$ times have been observed in $^{87}$Sr in tweezers, suggesting room for improvement in $^{171}$Yb.

Second, we note that a promising direction for future work is to store qubits in the metastable \tripletPzero\ state, which has a spontaneous emission-limited lifetime of $\tau = 20$ s. This preserves the benefits of the nuclear spin qubit in \singletS, but also enables single-photon Rydberg excitation ($\lambda = 302$ nm). Single-photon gates can be much faster, resulting in reduced sensitivity to Doppler shifts and laser phase noise \cite{deleseleuc2018,levine2018}, and also do not experience intermediate state scattering. Using single-photon Rydberg excitation from \tripletPzero, ground-Rydberg Bell states with a fidelity greater than 99\% have been recently demonstrated in $^{88}$Sr \cite{madjarov2020}, and this should translate directly to nuclear spin gates in the $^3$P$_0$ state of \ybOne.

Third, we note that this work demonstrates the first observation of the $^3$S$_1$ Rydberg series in \ybOne, and to the best of our knowledge, the first measurement of the Rydberg blockade in a hyperfine isotope of any AEA. While the Rydberg blockade for Yb \singletS series in \ybFour\ was predicted to be very small \cite{vaillant2012}, there have been no predictions of the \tripletS blockade strength in \ybFour\ or \ybOne. We recently demonstrated that the \tripletS series in \ybFour\ has a strong blockade \cite{Wilson2019, burgers2021}, and in this work, we have extended these results to \ybOne. The observed blockade strength has significant uncertainty (Appendix C), but appears to be larger than the blockade in the alkali atoms Rb or Cs at the same $n$, which may be the result of a hyperfine-induced F\"orster resonance \cite{robicheaux2019}. Confirming this result requires further spectroscopic measurements, particularly of the $^3$P$_J$ Rydberg states.

Lastly, while all of the demonstrated gate operations in this work are global, the extension to local operations is straightforward. Local single-qubit operations can be implemented using degenerate Raman transitions, while local two-qubit operations can be implemented by locally addressing one or both beams involved in the Rydberg excitation, or by locally addressing the 369 nm control beam to create local light shifts on $\ket{r}$ \cite{burgers2021}. The latter approach benefits from less required addressing power for high fidelity gates \cite{burgers2021}, and is also significantly insensitive to variations in the local intensity, as the differential light shift is negligible. It can also be used to control single-photon Rydberg excitation (\emph{e.g.}, gates on nuclear spin qubits in \tripletPzero).

\section{Conclusion}
In this work, we have demonstrated the first universal set of quantum gate operations for nuclear spin qubits in AEA tweezer arrays, using $^{171}$Yb. In particular, we implemented imaging, initialization, readout, one- and two-qubit gate operations, and coherent control of Rydberg excitation using a light shift on an ion core transition. In recent work, AEAs in optical tweezers have also been used to demonstrate extremely high fidelity atom detection \cite{covey2019}, highly coherent clock state manipulation \cite{Norcia2019}, record Rydberg entanglement fidelities \cite{madjarov2020} and efficient light shifting techniques and state detection using the ion core \cite{Lochead2013,madjarov2020,burgers2021}. Combining these techniques with nuclear spin qubits is a very promising avenue for realizing scalable, high-fidelity quantum gate operations, as well as entanglement-enhanced tweezer clocks using fermionic isotopes \cite{gil2014,kaubruegger2019}.

\section{Acknowledgments}
We gratefully acknowledge Sam Saskin for contributions to the experiment and developing an understanding of \ybOne\ Rydberg states, and Francesco Scazza for assistance with preliminary measurements of magic trapping wavelengths. This work was supported by ARO (W911NF-18-10215), ONR (N00014-20-1-2426), DARPA ONISQ (W911NF-20-10021) and the Sloan Foundation.

\emph{Note:} While completing this work, we became aware of complementary work demonstrating $^{171}$Yb qubits in tweezer arrays \cite{jenkins2021}.

\appendix
\section{Experimental apparatus}\label{app:exp}
A hot Yb atomic beam from an oven is cooled in a 2D magneto-optical trap (MOT) on the $^1$S$_0\rightarrow^1$P$_1$ transition ($\Gamma=2\pi\times31$ MHz, $\lambda=399$ nm). Cold atoms are pushed into a science chamber glass cell and loaded into a 3D MOT on the \tripletPone transition. Atoms are loaded from the MOT into tweezers with $\lambda=486.78$ nm (Coherent Genesis MX). The array is generated using a crossed pair of acousto-optic deflectors (AODs) controlled using an arbitrary waveform generator (AWG) from Spectrum Instrumentation (M4i.6622-x8), and projected through a 0.6 NA objective (Special Optics). Atoms are imaged using a single, retro-reflected beam with a projection onto all three trap axes. We do not apply any repumping light to depopulate the $^3$P$_{0,2}$ states. A detailed description of the apparatus can be found in Ref. \cite{Saskin2019}.

The Rydberg laser system consists of two sources: 556 nm (green, \singletS$\rightarrow$\tripletPone) and 308 nm (UV, \tripletPone$\rightarrow6s75s$\tripletS $F'=3/2, m_{F'}=3/2$) light. The green excitation light is generated via sum frequency generation in a PPLN crystal, using an amplified erbium fiber laser (1565 nm) and a titanium-sapphire (TiS) laser at 862 nm (M Squared Solstis). To produce UV light, we first generate 616 nm (orange) light using the same process described above, then double it in a resonant cavity (LEOS Rondo). The green and orange light frequencies are stabilized on a ULE reference cavity.

The green ($\sigma^+/\sigma^-$-polarized) and UV ($\pi$-polarized) beams are counter-propagating and focused to 30 $\mu$m and $15$ $\mu$m $1/e^2$ radius at the atoms, respectively. Typical measured single-photon Rabi frequencies are $\Omega_\mathrm{green}=2\pi\times 5.5$ MHz and $\Omega_{\mathrm{UV}}= 2\pi\times 5.5$ MHz. The latter value is obtained with a UV power of approximately $P_\mathrm{UV}=18$ mW. The 369 nm control beam used for autoionization and light shifts is co-propagating with the 308 nm Rydberg laser, but orthogonally polarized. The position and angle of the UV beam are monitored using cameras to ensure stable alignment to the atoms.

\section{\ybOne\ magic wavelength}\label{app:pol}
\begin{figure}[ht]
\centering
\includegraphics[width=1\linewidth]{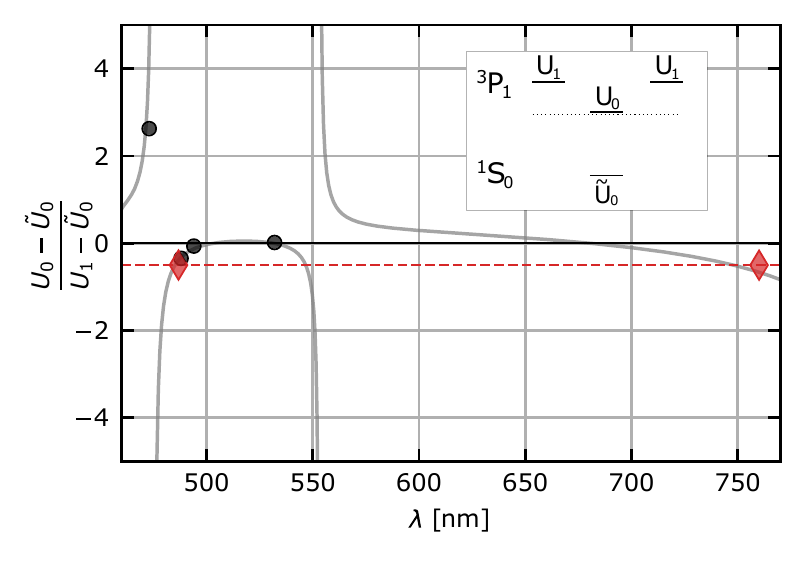}
\vspace{-5mm}
\caption{\label{fig:polarizability} Ratio of the differential light shifts for the \singletS-\tripletPone transition in $^{174}$Yb, for the $m_{J'}=0$ ($\Delta U_0 = U_0 - \tilde{U}_0$) and $|m_{J'}|=1$ ($\Delta U_1 = U_1 - \tilde{U}_0$) transitions. The black points show measurements from $^{174}$Yb, while the red points show inferred values from measurements with $^{171}$Yb. A magic wavelength for the $^{171}$Yb \singletS-\tripletPone $F'=3/2, |m_{F'}|=1/2$ transition occurs when $\Delta U_0/\Delta U_1 = -1/2$ (dashed red line), which we confirmed in \ybOne\ at 486.78 nm and near 759 nm. The gray line is an approximate model to guide the eye. (inset) Level diagram indicating the light shift on each state in \ybFour.}
\vspace{0mm}
\end{figure}

To implement narrow-line cooling and imaging in \ybOne, we have determined a new magic wavelength for the \singletS - \tripletPone transition in this isotope. We begin by measuring the differential light shifts for the $\pi$- and $\sigma$-polarized transitions in \ybFour, $\Delta U_0 = U_0 - \tilde{U}_0$ and $\Delta U_1 = U_1 - \tilde{U}_0$, respectively (Fig. \ref{fig:polarizability}). We report the ratio of these quantities, $\Delta U_0/\Delta U_1$, which is not subject to calibration errors in the absolute intensity. The data is qualitatively reproduced by a model incorporating computed matrix elements \cite{porsev1999}, with the inclusion of free parameters accounting for state-dependent core polarizability \cite{Arora2007}.

The magic wavelengths for \ybOne\ can be computed from the shifts of the \ybFour\ transitions. A magic wavelength for the transition to the $(F',|m_{F'}|) = (3/2,3/2)$ excited states occurs when $\Delta U_1 = 0$, while the transition to the $(F',|m_{F'}|) = (3/2,1/2)$ states is magic when $\Delta U_0/\Delta U_1 = -1/2$. We focus on the latter case because the dipole orientation provides more favorable fluorescence collection efficiency.

From the data in Fig. \ref{fig:polarizability}b, we predict that such a magic wavelength should occur near 486 nm and near 750 nm. Using direct measurements in \ybOne, we have determined a precise value for the short-wavelength magic wavelength (Fig. 1d). We have also experimentally confirmed that the \singletS - \tripletPzero clock transition magic wavelength at 759.35 nm \cite{barber2008,lemke2009} is also very close to magic for \singletS - \tripletPone, by measuring the light shift in a single tweezer at this wavelength.

\section{\ybOne\ Rydberg states}\label{app:blockade}
In this work, we use the Rydberg state $6s75s$ $^3$S$_1$, with $F'=3/2$. The Yb $^3$S$_1$ Rydberg series in \ybFour\ was first observed in Ref. \cite{Wilson2019}, and to the best of our knowledge, this series has not previously been studied in \ybOne. The Rydberg states of hyperfine isotopes in divalent atoms have a rich structure arising from the large hyperfine coupling in the ion core (12.6 GHz in the case of \ybOne) \cite{liao1980,beigang1981}. This gives rise to distinct Rydberg series converging to different ionization limits associated with different hyperfine states in the ion core. The emergence of these series from the $LS$-coupled states can be understood to arise from singlet-triplet mixing induced by hyperfine coupling \cite{liao1980,beigang1981}.

In this picture, the \tripletS\ $F'=3/2$ series is simple: it does not experience singlet-triplet mixing, and converges to the Yb$^+$ $F=1$ threshold. We identify this state by its quantum defect relative to this threshold, which is the same as the \ybFour\ \tripletS series ($\delta=4.439$), and also by the measured $g$ factor $g_F = 1.35$, consistent with the theoretical value of $4/3$. The total excitation energy from $^1$S$_0$ to $\ket{r}$ is 1511.569505(60) THz.

However, many of the other Rydberg series in \ybOne\ have more complex behavior, which makes calculations of the interaction potentials rather challenging \cite{robicheaux2019}. This challenge is compounded by the strongly perturbed $^3$P$_J$ series \cite{aymar1984}.

\begin{figure}[t]
\centering
\includegraphics[width=1\linewidth]{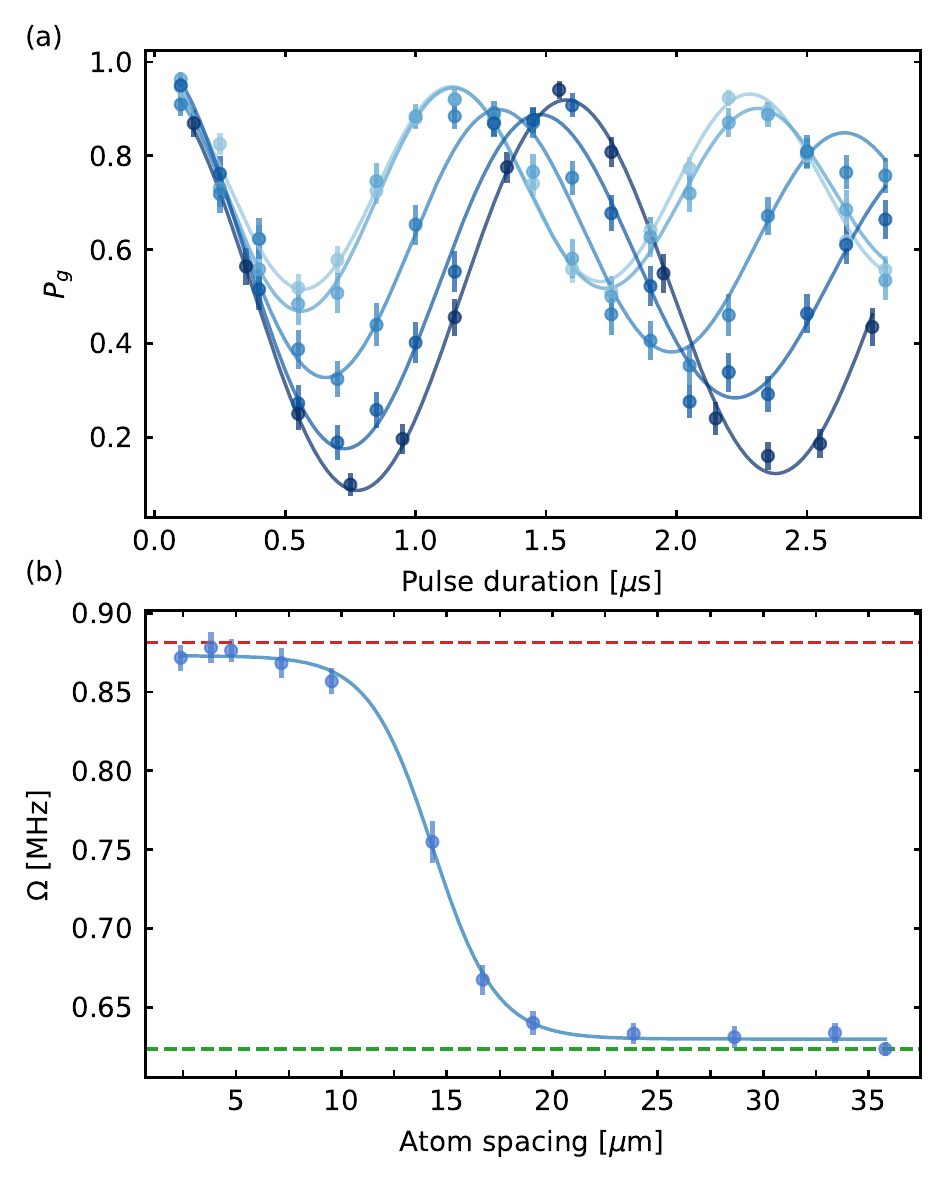}
\vspace{-5mm}
\caption{\label{fig:blockade} Measuring the blockade radius. (a) We observe Rabi oscillations dimers with atom separations between 36 $\mu$m (dark blue) and 2.4 $\mu$m (light blue). (b) The fitted oscillation frequency as a function of atomic spacing. The oscillation frequency increases from $\Omega_r=2\pi\times0.63$ MHz (green dashed line) to $\sqrt{2}\Omega_r$ (red dashed line) in the blockade regime. The line is a phenomenological fit to extract the blockade radius $R_b$, where $C_6/R_b^6 = \Omega_r$. This occurs when the oscillation frequency is halfway between $\Omega_r$ and $\sqrt{2}\Omega_r$.}
\vspace{0mm}
\end{figure}

Therefore, to validate the use of the Rydberg blockade in \ybOne, we have experimentally measured the van der Waals interaction for the $6s75s$ $^3$S$_1$ $\ket{F,m_F}=\ket{3/2,3/2}$ state (Fig. \ref{fig:blockade}). The measurement is performed by preparing atom pairs with varying spacing, driving the transition to $\ket{r}$ on resonance, and fitting the frequency of the observed Rabi oscillations at each separation. This frequency increases from $\Omega$ to $\sqrt{2} \Omega$ when $V > \Omega$.

We find a value for the blockade radius of $R_b=14(1.4)$ $\mu$m when $\Omega = 2 \pi \times 0.63$ MHz. This implies a $C_6$ coefficient of $5(3)$ THz$\cdot\mu$m$^6$. The large uncertainty in this value comes from uncertainty in the magnification of our imaging system. We have estimated the magnification from the component specifications, but not directly measured it, and assign an uncertainty of $\pm 10\%$.

We note the estimated $C_6$ is large, $2.7$ times larger than for S states in Rb at the same $n$ \cite{singer2005}. We have also measured $C_6$ for $n=50$ in \ybFour\ using the same technique and apparatus, finding $15(8)$ GHz$\cdot\mu$m$^6$ \cite{burgers2021}. Scaling to $n=75$ by $(n^*)^{11}$ gives a value 2.5 times smaller than the observed \ybOne\ $n=75$ value. Both observations are consistent with a smaller F\"orster energy defect \cite{saffman2010} giving rise to an enhanced blockade in \ybOne, and we note that hyperfine-induced F\"orster resonances in $^{87}$Sr were recently theoretically predicted \cite{robicheaux2019}. A detailed characterization of the \ybOne\ Rydberg series, as well as a detailed calculation of their interaction potentials and comparison with experiment, is left to future work.

\section{Gate fidelity calculations}\label{app:fidelity}
In this section, we estimate the two-qubit gate fidelity from the experimental data, incorporating the effects of Rydberg state leakage and SPAM errors. Our approach follows Ref. \cite{Levine2019} closely.

We wish to estimate the Bell state fidelity using the expression \cite{sackett2000}:
\begin{equation}
    \mathcal{F} = \frac{1}{2}(p_{00} + p_{11} + \mathcal{C}),
\end{equation} 
where $p_{00}$ ($p_{11}$) is the population in state $\ket{00}$ ($\ket{11}$) and $\mathcal{C}$ is the contrast obtained from the parity oscillation measurement (Fig. 4e).

In the experiment, population can also reside outside of the computational states $\{\ket{0}, \ket{1}\}$, for example, by leaking to $\ket{r}$. Non-computational states will not fluoresce during the state readout, and will therefore be mis-identified as being in $\ket{1}$ (Section \ref{sec:sq}). This leads to an overestimation of $p_{11}$. We can explicitly write the probability of the measurement outcomes $A_{ij}$ in terms of the atomic state populations as:

\begin{equation}\label{eqn:A}
\begin{split}
    A_{00} & = p_{00}, \\
    A_{01} & = p_{01} + p_{0r}, \\
    A_{10} & = p_{10} + p_{r0}, \\
    A_{11} & = p_{11} + p_{1r} + p_{r1} + p_{rr},
\end{split}
\end{equation}
\noindent Here, the subscript $r$ represents any state outside of the computational basis.

To obtain a better estimate of the populations $p_{ij}$, we repeat the experiment without the blowout. In this case, both $\ket{0}$ and $\ket{1}$ are bright, while everything outside the computational basis is dark. The probability of each outcome, $B_{ij}$, can be written in terms of $p_{ij}$ as:
\begin{equation}\label{eqn:B}
\begin{split}
    B_{bb} & = p_{00} + p_{01} + p_{10} + p_{11}, \\
    B_{bd} & = p_{0r} + p_{1r}, \\
    B_{db} & = p_{r0} + p_{r1}, \\
    B_{dd} & = p_{rr}. 
\end{split}
\end{equation}
Combining these equations, we derive expressions for $p_{ij}$,
\begin{align*}
    p_{00} &= A_{00}, \\
    p_{01} &= A_{01} - B_{bd} + p_{1r}, \\
    p_{10} &= A_{10} - B_{db} + p_{r1}, \\
    p_{11} &= A_{11} - B_{dd} - B_{bd} -     B_{db} + p_{0r} + p_{r0}.
\end{align*}
Since probabilities are non-negative, we can obtain lower bounds for $p_{ij}$:
\begin{align}
    p_{00} &= A_{00}, \\
    p_{01} &\geq A_{01} - B_{bd}, \\
    p_{10} &\geq A_{10} - B_{db}, \\
    p_{11} &\geq A_{11} - (1 - B_{bb}). \label{eqn:p11}
\end{align}

Experimental measurements of $A_{ij}$ and $B_{ij}$ after preparing the Bell state are presented in Table \ref{tab:AnB}. The contrast of parity oscillation is $\mathcal{C}=0.79(3)$.

The raw Bell state fidelity computed from the $A_{ij}$ is:
\begin{equation}
\mathcal{F}_{\mathrm{raw}} = \frac{1}{2}(A_{00} + A_{11} + \mathcal{C}) = 85(2)\%.
\end{equation}

Equation (\ref{eqn:p11}) yields a lower bound for $p_{11} \geq 0.40(2)$, while $p_{00}=A_{00}$ and $\mathcal{C}$ remain unchanged. This gives rise to a lower bound on the true Bell state fidelity of:
$$
\mathcal{F} = \frac{1}{2}(p_{00} + p_{11} + \mathcal{C}) \geq 80(2)\%.
$$
\begin{table}[tp]
    \centering
    \begin{tabular}{|c|c||c|c|}
    \hline
        $A_{00}$ &  0.41(2) & $B_{bb}$ &  0.903(13)\\
        $A_{01}$ &  0.046(8) & $B_{bd}$ &  0.044(9)\\
        $A_{10}$ &  0.047(8) & $B_{db}$ &  0.053(10)\\
        $A_{11}$ &  0.50(2) & $B_{dd}$ &  $<0.01$\\
    \hline
    \end{tabular}
    \caption{Experimentally measured $A_{ij}$ and $B_{ij}$ in equations (\ref{eqn:A}) and (\ref{eqn:B}).}
    \label{tab:AnB}
\end{table}
This is lower than $\mathcal{F}_{\mathrm{raw}}$ because it accounts for leakage out of the computational space, which has a greater impact on the fidelity than population in the wrong computational state.

Additionally, we wish to estimate the impact state preparation and measurement (SPAM) errors. In our experiment, the main source of SPAM error is atom loss during the initialization and measurement phases.

If one of the two atoms is lost before the gate, the final state of the remaining atom (for the ideal circuit) is $\ket{\psi} = (\ket{0} + e^{i\phi}\ket{1})/\sqrt{2}$, which yields the measurement result $11$ or $10/01$ with equal probability. These events do not contribute to the parity oscillation, since the oscillation period is different from the Bell state.

Meanwhile, if an atom is lost after the gate, the remaining atom can be described by the mixed state $\rho = \mathbb{1}/2$. This gives the same measurement results as if the atom was lost before the gate, and does not oscillate at all during the parity oscillation measurement.

Therefore, atom loss before and after the gate contribute in the same way to the measurement results. Given an overall loss rate of $\epsilon$ per atom, we can correct $p_{00}$ and $\mathcal{C}$ by multiplying by a factor of $(1-\epsilon)^{-2}$,
\begin{align}
    p_{00}^c &= \frac{p_{00}}{(1-\epsilon)^2},  \label{eqn:p00c}\\
    \mathcal{C}^c &= \frac{\mathcal{C}}{(1-\epsilon)^2} \label{eqn:Cc}.
\end{align}
In the case of $p_{11}$, we must consider that losing one of the atoms results in the outcome $11$ with probability $1/2$, while losing both the atoms results in $11$ with probability $1$. Thus, we can write
\begin{align*}
    p_{11} &= p_{11}^c (1-\epsilon)^2 + 0.5\times\epsilon(1-\epsilon) + 0.5\times(1-\epsilon)\epsilon + \epsilon^2, \\
    & = p_{11}^c (1-\epsilon)^2 + \epsilon.
\end{align*}
This gives rise to the corrected population in $\ket{11}$,
\begin{equation}
    p_{11}^c = \frac{p_{11} - \epsilon}{(1-\epsilon)^2}. \label{eqn:p11c}
\end{equation}
Using independent experiments, we conservatively estimate the atom loss probability to be $\epsilon \geq 2.4(6)\%$. Utilizing equations (\ref{eqn:p00c}), (\ref{eqn:Cc}) and (\ref{eqn:p11c}), we are able to put a lower bound on the Bell state fidelity after SPAM correction:

\begin{equation}
\mathcal{F}^{c} = \frac{1}{2}(p_{00}^c + p_{11}^c + \mathcal{C}^c) \geq 83(2)\%.
\end{equation}

In section \ref{sec:ls}, we demonstrate that the gate can also be suppressed by the 369 nm laser. Therefore, the final state should be $\ket{00}$ (see Fig \ref{fig:autionization}a). In the experiment, we measure a raw fidelity of $\mathcal{F}_I = p_{00} = 84(2)\%$. On the day that these measurements were performed, the atom survival rate after imaging and rearrangement was worse, and we independently measured the atom loss $\epsilon = 4.6(5)\%$. From this measurement, we correct the fidelity $\mathcal{F}^c_I = p^c_{00} = 93(2)\%$ using equation (\ref{eqn:p00c}).

\bibliography{171Yb.bib}

\begin{thebibliography}{53}%
\makeatletter
\providecommand \@ifxundefined [1]{%
 \@ifx{#1\undefined}
}%
\providecommand \@ifnum [1]{%
 \ifnum #1\expandafter \@firstoftwo
 \else \expandafter \@secondoftwo
 \fi
}%
\providecommand \@ifx [1]{%
 \ifx #1\expandafter \@firstoftwo
 \else \expandafter \@secondoftwo
 \fi
}%
\providecommand \natexlab [1]{#1}%
\providecommand \enquote  [1]{``#1''}%
\providecommand \bibnamefont  [1]{#1}%
\providecommand \bibfnamefont [1]{#1}%
\providecommand \citenamefont [1]{#1}%
\providecommand \href@noop [0]{\@secondoftwo}%
\providecommand \href [0]{\begingroup \@sanitize@url \@href}%
\providecommand \@href[1]{\@@startlink{#1}\@@href}%
\providecommand \@@href[1]{\endgroup#1\@@endlink}%
\providecommand \@sanitize@url [0]{\catcode `\\12\catcode `\$12\catcode
  `\&12\catcode `\#12\catcode `\^12\catcode `\_12\catcode `\%12\relax}%
\providecommand \@@startlink[1]{}%
\providecommand \@@endlink[0]{}%
\providecommand \url  [0]{\begingroup\@sanitize@url \@url }%
\providecommand \@url [1]{\endgroup\@href {#1}{\urlprefix }}%
\providecommand \urlprefix  [0]{URL }%
\providecommand \Eprint [0]{\href }%
\providecommand \doibase [0]{https://doi.org/}%
\providecommand \selectlanguage [0]{\@gobble}%
\providecommand \bibinfo  [0]{\@secondoftwo}%
\providecommand \bibfield  [0]{\@secondoftwo}%
\providecommand \translation [1]{[#1]}%
\providecommand \BibitemOpen [0]{}%
\providecommand \bibitemStop [0]{}%
\providecommand \bibitemNoStop [0]{.\EOS\space}%
\providecommand \EOS [0]{\spacefactor3000\relax}%
\providecommand \BibitemShut  [1]{\csname bibitem#1\endcsname}%
\let\auto@bib@innerbib\@empty
\bibitem [{\citenamefont {Saffman}(2016)}]{saffman2016}%
  \BibitemOpen
  \bibfield  {author} {\bibinfo {author} {\bibfnamefont {M.}~\bibnamefont
  {Saffman}},\ }\href {https://doi.org/10.1088/0953-4075/49/20/202001}
  {\bibfield  {journal} {\bibinfo  {journal} {Journal of Physics B: Atomic,
  Molecular and Optical Physics}\ }\textbf {\bibinfo {volume} {49}},\ \bibinfo
  {pages} {202001} (\bibinfo {year} {2016})}\BibitemShut {NoStop}%
\bibitem [{\citenamefont {Browaeys}\ and\ \citenamefont
  {Lahaye}(2020)}]{browaeys2020}%
  \BibitemOpen
  \bibfield  {author} {\bibinfo {author} {\bibfnamefont {A.}~\bibnamefont
  {Browaeys}}\ and\ \bibinfo {author} {\bibfnamefont {T.}~\bibnamefont
  {Lahaye}},\ }\href {https://doi.org/10.1038/s41567-019-0733-z} {\bibfield
  {journal} {\bibinfo  {journal} {Nature Physics}\ }\textbf {\bibinfo {volume}
  {16}},\ \bibinfo {pages} {132} (\bibinfo {year} {2020})}\BibitemShut
  {NoStop}%
\bibitem [{\citenamefont {Ebadi}\ \emph {et~al.}(2021)\citenamefont {Ebadi},
  \citenamefont {Wang}, \citenamefont {Levine}, \citenamefont {Keesling},
  \citenamefont {Semeghini}, \citenamefont {Omran}, \citenamefont {Bluvstein},
  \citenamefont {Samajdar}, \citenamefont {Pichler}, \citenamefont {Ho},
  \citenamefont {Choi}, \citenamefont {Sachdev}, \citenamefont {Greiner},
  \citenamefont {Vuleti{\'c}},\ and\ \citenamefont {Lukin}}]{ebadi2020}%
  \BibitemOpen
  \bibfield  {author} {\bibinfo {author} {\bibfnamefont {S.}~\bibnamefont
  {Ebadi}}, \bibinfo {author} {\bibfnamefont {T.~T.}\ \bibnamefont {Wang}},
  \bibinfo {author} {\bibfnamefont {H.}~\bibnamefont {Levine}}, \bibinfo
  {author} {\bibfnamefont {A.}~\bibnamefont {Keesling}}, \bibinfo {author}
  {\bibfnamefont {G.}~\bibnamefont {Semeghini}}, \bibinfo {author}
  {\bibfnamefont {A.}~\bibnamefont {Omran}}, \bibinfo {author} {\bibfnamefont
  {D.}~\bibnamefont {Bluvstein}}, \bibinfo {author} {\bibfnamefont
  {R.}~\bibnamefont {Samajdar}}, \bibinfo {author} {\bibfnamefont
  {H.}~\bibnamefont {Pichler}}, \bibinfo {author} {\bibfnamefont {W.~W.}\
  \bibnamefont {Ho}}, \bibinfo {author} {\bibfnamefont {S.}~\bibnamefont
  {Choi}}, \bibinfo {author} {\bibfnamefont {S.}~\bibnamefont {Sachdev}},
  \bibinfo {author} {\bibfnamefont {M.}~\bibnamefont {Greiner}}, \bibinfo
  {author} {\bibfnamefont {V.}~\bibnamefont {Vuleti{\'c}}},\ and\ \bibinfo
  {author} {\bibfnamefont {M.~D.}\ \bibnamefont {Lukin}},\ }\href
  {https://doi.org/10.1038/s41586-021-03582-4} {\bibfield  {journal} {\bibinfo
  {journal} {Nature}\ }\textbf {\bibinfo {volume} {595}},\ \bibinfo {pages}
  {227} (\bibinfo {year} {2021})}\BibitemShut {NoStop}%
\bibitem [{\citenamefont {Scholl}\ \emph {et~al.}(2021)\citenamefont {Scholl},
  \citenamefont {Schuler}, \citenamefont {Williams}, \citenamefont
  {Eberharter}, \citenamefont {Barredo}, \citenamefont {Schymik}, \citenamefont
  {Lienhard}, \citenamefont {Henry}, \citenamefont {Lang}, \citenamefont
  {Lahaye}, \citenamefont {L{\"a}uchli},\ and\ \citenamefont
  {Browaeys}}]{scholl2020}%
  \BibitemOpen
  \bibfield  {author} {\bibinfo {author} {\bibfnamefont {P.}~\bibnamefont
  {Scholl}}, \bibinfo {author} {\bibfnamefont {M.}~\bibnamefont {Schuler}},
  \bibinfo {author} {\bibfnamefont {H.~J.}\ \bibnamefont {Williams}}, \bibinfo
  {author} {\bibfnamefont {A.~A.}\ \bibnamefont {Eberharter}}, \bibinfo
  {author} {\bibfnamefont {D.}~\bibnamefont {Barredo}}, \bibinfo {author}
  {\bibfnamefont {K.-N.}\ \bibnamefont {Schymik}}, \bibinfo {author}
  {\bibfnamefont {V.}~\bibnamefont {Lienhard}}, \bibinfo {author}
  {\bibfnamefont {L.-P.}\ \bibnamefont {Henry}}, \bibinfo {author}
  {\bibfnamefont {T.~C.}\ \bibnamefont {Lang}}, \bibinfo {author}
  {\bibfnamefont {T.}~\bibnamefont {Lahaye}}, \bibinfo {author} {\bibfnamefont
  {A.~M.}\ \bibnamefont {L{\"a}uchli}},\ and\ \bibinfo {author} {\bibfnamefont
  {A.}~\bibnamefont {Browaeys}},\ }\href
  {https://doi.org/10.1038/s41586-021-03585-1} {\bibfield  {journal} {\bibinfo
  {journal} {Nature}\ }\textbf {\bibinfo {volume} {595}},\ \bibinfo {pages}
  {233} (\bibinfo {year} {2021})}\BibitemShut {NoStop}%
\bibitem [{\citenamefont {Semeghini}\ \emph {et~al.}(2021)\citenamefont
  {Semeghini}, \citenamefont {Levine}, \citenamefont {Keesling}, \citenamefont
  {Ebadi}, \citenamefont {Wang}, \citenamefont {Bluvstein}, \citenamefont
  {Verresen}, \citenamefont {Pichler}, \citenamefont {Kalinowski},
  \citenamefont {Samajdar}, \citenamefont {Omran}, \citenamefont {Sachdev},
  \citenamefont {Vishwanath}, \citenamefont {Greiner}, \citenamefont
  {Vuletić},\ and\ \citenamefont {Lukin}}]{semeghini2021}%
  \BibitemOpen
  \bibfield  {author} {\bibinfo {author} {\bibfnamefont {G.}~\bibnamefont
  {Semeghini}}, \bibinfo {author} {\bibfnamefont {H.}~\bibnamefont {Levine}},
  \bibinfo {author} {\bibfnamefont {A.}~\bibnamefont {Keesling}}, \bibinfo
  {author} {\bibfnamefont {S.}~\bibnamefont {Ebadi}}, \bibinfo {author}
  {\bibfnamefont {T.~T.}\ \bibnamefont {Wang}}, \bibinfo {author}
  {\bibfnamefont {D.}~\bibnamefont {Bluvstein}}, \bibinfo {author}
  {\bibfnamefont {R.}~\bibnamefont {Verresen}}, \bibinfo {author}
  {\bibfnamefont {H.}~\bibnamefont {Pichler}}, \bibinfo {author} {\bibfnamefont
  {M.}~\bibnamefont {Kalinowski}}, \bibinfo {author} {\bibfnamefont
  {R.}~\bibnamefont {Samajdar}}, \bibinfo {author} {\bibfnamefont
  {A.}~\bibnamefont {Omran}}, \bibinfo {author} {\bibfnamefont
  {S.}~\bibnamefont {Sachdev}}, \bibinfo {author} {\bibfnamefont
  {A.}~\bibnamefont {Vishwanath}}, \bibinfo {author} {\bibfnamefont
  {M.}~\bibnamefont {Greiner}}, \bibinfo {author} {\bibfnamefont
  {V.}~\bibnamefont {Vuletić}},\ and\ \bibinfo {author} {\bibfnamefont
  {M.~D.}\ \bibnamefont {Lukin}},\ }\href
  {https://doi.org/10.1126/science.abi8794} {\bibfield  {journal} {\bibinfo
  {journal} {Science}\ }\textbf {\bibinfo {volume} {374}},\ \bibinfo {pages}
  {1242} (\bibinfo {year} {2021})}\BibitemShut {NoStop}%
\bibitem [{\citenamefont {Graham}\ \emph {et~al.}(2019)\citenamefont {Graham},
  \citenamefont {Kwon}, \citenamefont {Grinkemeyer}, \citenamefont {Marra},
  \citenamefont {Jiang}, \citenamefont {Lichtman}, \citenamefont {Sun},
  \citenamefont {Ebert},\ and\ \citenamefont {Saffman}}]{graham2019}%
  \BibitemOpen
  \bibfield  {author} {\bibinfo {author} {\bibfnamefont {T.~M.}\ \bibnamefont
  {Graham}}, \bibinfo {author} {\bibfnamefont {M.}~\bibnamefont {Kwon}},
  \bibinfo {author} {\bibfnamefont {B.}~\bibnamefont {Grinkemeyer}}, \bibinfo
  {author} {\bibfnamefont {Z.}~\bibnamefont {Marra}}, \bibinfo {author}
  {\bibfnamefont {X.}~\bibnamefont {Jiang}}, \bibinfo {author} {\bibfnamefont
  {M.~T.}\ \bibnamefont {Lichtman}}, \bibinfo {author} {\bibfnamefont
  {Y.}~\bibnamefont {Sun}}, \bibinfo {author} {\bibfnamefont {M.}~\bibnamefont
  {Ebert}},\ and\ \bibinfo {author} {\bibfnamefont {M.}~\bibnamefont
  {Saffman}},\ }\href {https://doi.org/10.1103/PhysRevLett.123.230501}
  {\bibfield  {journal} {\bibinfo  {journal} {Phys. Rev. Lett.}\ }\textbf
  {\bibinfo {volume} {123}},\ \bibinfo {pages} {230501} (\bibinfo {year}
  {2019})}\BibitemShut {NoStop}%
\bibitem [{\citenamefont {Levine}\ \emph {et~al.}(2019)\citenamefont {Levine},
  \citenamefont {Keesling}, \citenamefont {Semeghini}, \citenamefont {Omran},
  \citenamefont {Wang}, \citenamefont {Ebadi}, \citenamefont {Bernien},
  \citenamefont {Greiner}, \citenamefont {Vuleti\ifmmode~\acute{c}\else
  \'{c}\fi{}}, \citenamefont {Pichler},\ and\ \citenamefont
  {Lukin}}]{Levine2019}%
  \BibitemOpen
  \bibfield  {author} {\bibinfo {author} {\bibfnamefont {H.}~\bibnamefont
  {Levine}}, \bibinfo {author} {\bibfnamefont {A.}~\bibnamefont {Keesling}},
  \bibinfo {author} {\bibfnamefont {G.}~\bibnamefont {Semeghini}}, \bibinfo
  {author} {\bibfnamefont {A.}~\bibnamefont {Omran}}, \bibinfo {author}
  {\bibfnamefont {T.~T.}\ \bibnamefont {Wang}}, \bibinfo {author}
  {\bibfnamefont {S.}~\bibnamefont {Ebadi}}, \bibinfo {author} {\bibfnamefont
  {H.}~\bibnamefont {Bernien}}, \bibinfo {author} {\bibfnamefont
  {M.}~\bibnamefont {Greiner}}, \bibinfo {author} {\bibfnamefont
  {V.}~\bibnamefont {Vuleti\ifmmode~\acute{c}\else \'{c}\fi{}}}, \bibinfo
  {author} {\bibfnamefont {H.}~\bibnamefont {Pichler}},\ and\ \bibinfo {author}
  {\bibfnamefont {M.~D.}\ \bibnamefont {Lukin}},\ }\href
  {https://doi.org/10.1103/PhysRevLett.123.170503} {\bibfield  {journal}
  {\bibinfo  {journal} {Phys. Rev. Lett.}\ }\textbf {\bibinfo {volume} {123}},\
  \bibinfo {pages} {170503} (\bibinfo {year} {2019})}\BibitemShut {NoStop}%
\bibitem [{\citenamefont {Omran}\ \emph {et~al.}(2019)\citenamefont {Omran},
  \citenamefont {Levine}, \citenamefont {Keesling}, \citenamefont {Semeghini},
  \citenamefont {Wang}, \citenamefont {Ebadi}, \citenamefont {Bernien},
  \citenamefont {Zibrov}, \citenamefont {Pichler}, \citenamefont {Choi},
  \citenamefont {Cui}, \citenamefont {Rossignolo}, \citenamefont {Rembold},
  \citenamefont {Montangero}, \citenamefont {Calarco}, \citenamefont {Endres},
  \citenamefont {Greiner}, \citenamefont {Vuleti{\'c}},\ and\ \citenamefont
  {Lukin}}]{omran2019}%
  \BibitemOpen
  \bibfield  {author} {\bibinfo {author} {\bibfnamefont {A.}~\bibnamefont
  {Omran}}, \bibinfo {author} {\bibfnamefont {H.}~\bibnamefont {Levine}},
  \bibinfo {author} {\bibfnamefont {A.}~\bibnamefont {Keesling}}, \bibinfo
  {author} {\bibfnamefont {G.}~\bibnamefont {Semeghini}}, \bibinfo {author}
  {\bibfnamefont {T.~T.}\ \bibnamefont {Wang}}, \bibinfo {author}
  {\bibfnamefont {S.}~\bibnamefont {Ebadi}}, \bibinfo {author} {\bibfnamefont
  {H.}~\bibnamefont {Bernien}}, \bibinfo {author} {\bibfnamefont {A.~S.}\
  \bibnamefont {Zibrov}}, \bibinfo {author} {\bibfnamefont {H.}~\bibnamefont
  {Pichler}}, \bibinfo {author} {\bibfnamefont {S.}~\bibnamefont {Choi}},
  \bibinfo {author} {\bibfnamefont {J.}~\bibnamefont {Cui}}, \bibinfo {author}
  {\bibfnamefont {M.}~\bibnamefont {Rossignolo}}, \bibinfo {author}
  {\bibfnamefont {P.}~\bibnamefont {Rembold}}, \bibinfo {author} {\bibfnamefont
  {S.}~\bibnamefont {Montangero}}, \bibinfo {author} {\bibfnamefont
  {T.}~\bibnamefont {Calarco}}, \bibinfo {author} {\bibfnamefont
  {M.}~\bibnamefont {Endres}}, \bibinfo {author} {\bibfnamefont
  {M.}~\bibnamefont {Greiner}}, \bibinfo {author} {\bibfnamefont
  {V.}~\bibnamefont {Vuleti{\'c}}},\ and\ \bibinfo {author} {\bibfnamefont
  {M.~D.}\ \bibnamefont {Lukin}},\ }\href
  {https://doi.org/10.1126/science.aax9743} {\bibfield  {journal} {\bibinfo
  {journal} {Science}\ }\textbf {\bibinfo {volume} {365}},\ \bibinfo {pages}
  {570} (\bibinfo {year} {2019})}\BibitemShut {NoStop}%
\bibitem [{\citenamefont {Choi}\ \emph {et~al.}(2021)\citenamefont {Choi},
  \citenamefont {Shaw}, \citenamefont {Madjarov}, \citenamefont {Xie},
  \citenamefont {Covey}, \citenamefont {Cotler}, \citenamefont {Mark},
  \citenamefont {Huang}, \citenamefont {Kale}, \citenamefont {Pichler},
  \citenamefont {Brandão}, \citenamefont {Choi},\ and\ \citenamefont
  {Endres}}]{choi2021}%
  \BibitemOpen
  \bibfield  {author} {\bibinfo {author} {\bibfnamefont {J.}~\bibnamefont
  {Choi}}, \bibinfo {author} {\bibfnamefont {A.~L.}\ \bibnamefont {Shaw}},
  \bibinfo {author} {\bibfnamefont {I.~S.}\ \bibnamefont {Madjarov}}, \bibinfo
  {author} {\bibfnamefont {X.}~\bibnamefont {Xie}}, \bibinfo {author}
  {\bibfnamefont {J.~P.}\ \bibnamefont {Covey}}, \bibinfo {author}
  {\bibfnamefont {J.~S.}\ \bibnamefont {Cotler}}, \bibinfo {author}
  {\bibfnamefont {D.~K.}\ \bibnamefont {Mark}}, \bibinfo {author}
  {\bibfnamefont {H.-Y.}\ \bibnamefont {Huang}}, \bibinfo {author}
  {\bibfnamefont {A.}~\bibnamefont {Kale}}, \bibinfo {author} {\bibfnamefont
  {H.}~\bibnamefont {Pichler}}, \bibinfo {author} {\bibfnamefont {F.~G. S.~L.}\
  \bibnamefont {Brandão}}, \bibinfo {author} {\bibfnamefont {S.}~\bibnamefont
  {Choi}},\ and\ \bibinfo {author} {\bibfnamefont {M.}~\bibnamefont {Endres}},\
  }\href@noop {} {\bibinfo {title} {Emergent randomness and benchmarking from
  many-body quantum chaos}} (\bibinfo {year} {2021}),\ \Eprint
  {https://arxiv.org/abs/2103.03535} {arXiv:2103.03535 [quant-ph]} \BibitemShut
  {NoStop}%
\bibitem [{\citenamefont {Cooper}\ \emph {et~al.}(2018)\citenamefont {Cooper},
  \citenamefont {Covey}, \citenamefont {Madjarov}, \citenamefont {Porsev},
  \citenamefont {Safronova},\ and\ \citenamefont {Endres}}]{Cooper2018}%
  \BibitemOpen
  \bibfield  {author} {\bibinfo {author} {\bibfnamefont {A.}~\bibnamefont
  {Cooper}}, \bibinfo {author} {\bibfnamefont {J.~P.}\ \bibnamefont {Covey}},
  \bibinfo {author} {\bibfnamefont {I.~S.}\ \bibnamefont {Madjarov}}, \bibinfo
  {author} {\bibfnamefont {S.~G.}\ \bibnamefont {Porsev}}, \bibinfo {author}
  {\bibfnamefont {M.~S.}\ \bibnamefont {Safronova}},\ and\ \bibinfo {author}
  {\bibfnamefont {M.}~\bibnamefont {Endres}},\ }\href
  {https://doi.org/10.1103/PhysRevX.8.041055} {\bibfield  {journal} {\bibinfo
  {journal} {Phys. Rev. X}\ }\textbf {\bibinfo {volume} {8}},\ \bibinfo {pages}
  {041055} (\bibinfo {year} {2018})}\BibitemShut {NoStop}%
\bibitem [{\citenamefont {Norcia}\ \emph {et~al.}(2018)\citenamefont {Norcia},
  \citenamefont {Young},\ and\ \citenamefont {Kaufman}}]{Norcia2018}%
  \BibitemOpen
  \bibfield  {author} {\bibinfo {author} {\bibfnamefont {M.~A.}\ \bibnamefont
  {Norcia}}, \bibinfo {author} {\bibfnamefont {A.~W.}\ \bibnamefont {Young}},\
  and\ \bibinfo {author} {\bibfnamefont {A.~M.}\ \bibnamefont {Kaufman}},\
  }\href {https://doi.org/10.1103/PhysRevX.8.041054} {\bibfield  {journal}
  {\bibinfo  {journal} {Phys. Rev. X}\ }\textbf {\bibinfo {volume} {8}},\
  \bibinfo {pages} {041054} (\bibinfo {year} {2018})}\BibitemShut {NoStop}%
\bibitem [{\citenamefont {Barnes}\ \emph {et~al.}(2021)\citenamefont {Barnes},
  \citenamefont {Battaglino}, \citenamefont {Bloom}, \citenamefont {Cassella},
  \citenamefont {Coxe}, \citenamefont {Crisosto}, \citenamefont {King},
  \citenamefont {Kondov}, \citenamefont {Kotru}, \citenamefont {Larsen},
  \citenamefont {Lauigan}, \citenamefont {Lester}, \citenamefont {McDonald},
  \citenamefont {Megidish}, \citenamefont {Narayanaswami}, \citenamefont
  {Nishiguchi}, \citenamefont {Notermans}, \citenamefont {Peng}, \citenamefont
  {Ryou}, \citenamefont {Wu},\ and\ \citenamefont {Yarwood}}]{barnes2021}%
  \BibitemOpen
  \bibfield  {author} {\bibinfo {author} {\bibfnamefont {K.}~\bibnamefont
  {Barnes}}, \bibinfo {author} {\bibfnamefont {P.}~\bibnamefont {Battaglino}},
  \bibinfo {author} {\bibfnamefont {B.~J.}\ \bibnamefont {Bloom}}, \bibinfo
  {author} {\bibfnamefont {K.}~\bibnamefont {Cassella}}, \bibinfo {author}
  {\bibfnamefont {R.}~\bibnamefont {Coxe}}, \bibinfo {author} {\bibfnamefont
  {N.}~\bibnamefont {Crisosto}}, \bibinfo {author} {\bibfnamefont {J.~P.}\
  \bibnamefont {King}}, \bibinfo {author} {\bibfnamefont {S.~S.}\ \bibnamefont
  {Kondov}}, \bibinfo {author} {\bibfnamefont {K.}~\bibnamefont {Kotru}},
  \bibinfo {author} {\bibfnamefont {S.~C.}\ \bibnamefont {Larsen}}, \bibinfo
  {author} {\bibfnamefont {J.}~\bibnamefont {Lauigan}}, \bibinfo {author}
  {\bibfnamefont {B.~J.}\ \bibnamefont {Lester}}, \bibinfo {author}
  {\bibfnamefont {M.}~\bibnamefont {McDonald}}, \bibinfo {author}
  {\bibfnamefont {E.}~\bibnamefont {Megidish}}, \bibinfo {author}
  {\bibfnamefont {S.}~\bibnamefont {Narayanaswami}}, \bibinfo {author}
  {\bibfnamefont {C.}~\bibnamefont {Nishiguchi}}, \bibinfo {author}
  {\bibfnamefont {R.}~\bibnamefont {Notermans}}, \bibinfo {author}
  {\bibfnamefont {L.~S.}\ \bibnamefont {Peng}}, \bibinfo {author}
  {\bibfnamefont {A.}~\bibnamefont {Ryou}}, \bibinfo {author} {\bibfnamefont
  {T.-Y.}\ \bibnamefont {Wu}},\ and\ \bibinfo {author} {\bibfnamefont
  {M.}~\bibnamefont {Yarwood}},\ }\href@noop {} {\bibinfo {title} {Assembly and
  coherent control of a register of nuclear spin qubits}} (\bibinfo {year}
  {2021}),\ \Eprint {https://arxiv.org/abs/2108.04790} {arXiv:2108.04790
  [quant-ph]} \BibitemShut {NoStop}%
\bibitem [{\citenamefont {Saskin}\ \emph {et~al.}(2019)\citenamefont {Saskin},
  \citenamefont {Wilson}, \citenamefont {Grinkemeyer},\ and\ \citenamefont
  {Thompson}}]{Saskin2019}%
  \BibitemOpen
  \bibfield  {author} {\bibinfo {author} {\bibfnamefont {S.}~\bibnamefont
  {Saskin}}, \bibinfo {author} {\bibfnamefont {J.~T.}\ \bibnamefont {Wilson}},
  \bibinfo {author} {\bibfnamefont {B.}~\bibnamefont {Grinkemeyer}},\ and\
  \bibinfo {author} {\bibfnamefont {J.~D.}\ \bibnamefont {Thompson}},\ }\href
  {https://doi.org/10.1103/PhysRevLett.122.143002} {\bibfield  {journal}
  {\bibinfo  {journal} {Phys. Rev. Lett.}\ }\textbf {\bibinfo {volume} {122}},\
  \bibinfo {pages} {143002} (\bibinfo {year} {2019})}\BibitemShut {NoStop}%
\bibitem [{\citenamefont {Wilson}\ \emph {et~al.}(2019)\citenamefont {Wilson},
  \citenamefont {Saskin}, \citenamefont {Meng}, \citenamefont {Ma},
  \citenamefont {Dilip}, \citenamefont {Burgers},\ and\ \citenamefont
  {Thompson}}]{Wilson2019}%
  \BibitemOpen
  \bibfield  {author} {\bibinfo {author} {\bibfnamefont {J.}~\bibnamefont
  {Wilson}}, \bibinfo {author} {\bibfnamefont {S.}~\bibnamefont {Saskin}},
  \bibinfo {author} {\bibfnamefont {Y.}~\bibnamefont {Meng}}, \bibinfo {author}
  {\bibfnamefont {S.}~\bibnamefont {Ma}}, \bibinfo {author} {\bibfnamefont
  {R.}~\bibnamefont {Dilip}}, \bibinfo {author} {\bibfnamefont {A.~P.}\
  \bibnamefont {Burgers}},\ and\ \bibinfo {author} {\bibfnamefont {J.~D.}\
  \bibnamefont {Thompson}},\ }\href@noop {} {\bibinfo {title} {Trapped arrays
  of alkaline earth rydberg atoms in optical tweezers}} (\bibinfo {year}
  {2019}),\ \Eprint {https://arxiv.org/abs/1912.08754} {arXiv:1912.08754
  [quant-ph]} \BibitemShut {NoStop}%
\bibitem [{\citenamefont {Madjarov}\ \emph {et~al.}(2020)\citenamefont
  {Madjarov}, \citenamefont {Covey}, \citenamefont {Shaw}, \citenamefont
  {Choi}, \citenamefont {Kale}, \citenamefont {Cooper}, \citenamefont
  {Pichler}, \citenamefont {Schkolnik}, \citenamefont {Williams},\ and\
  \citenamefont {Endres}}]{madjarov2020}%
  \BibitemOpen
  \bibfield  {author} {\bibinfo {author} {\bibfnamefont {I.~S.}\ \bibnamefont
  {Madjarov}}, \bibinfo {author} {\bibfnamefont {J.~P.}\ \bibnamefont {Covey}},
  \bibinfo {author} {\bibfnamefont {A.~L.}\ \bibnamefont {Shaw}}, \bibinfo
  {author} {\bibfnamefont {J.}~\bibnamefont {Choi}}, \bibinfo {author}
  {\bibfnamefont {A.}~\bibnamefont {Kale}}, \bibinfo {author} {\bibfnamefont
  {A.}~\bibnamefont {Cooper}}, \bibinfo {author} {\bibfnamefont
  {H.}~\bibnamefont {Pichler}}, \bibinfo {author} {\bibfnamefont
  {V.}~\bibnamefont {Schkolnik}}, \bibinfo {author} {\bibfnamefont {J.~R.}\
  \bibnamefont {Williams}},\ and\ \bibinfo {author} {\bibfnamefont
  {M.}~\bibnamefont {Endres}},\ }\href
  {https://doi.org/10.1038/s41567-020-0903-z} {\bibfield  {journal} {\bibinfo
  {journal} {Nature Physics}\ }\textbf {\bibinfo {volume} {16}},\ \bibinfo
  {pages} {857} (\bibinfo {year} {2020})}\BibitemShut {NoStop}%
\bibitem [{\citenamefont {Burgers}\ \emph {et~al.}(2021)\citenamefont
  {Burgers}, \citenamefont {Ma}, \citenamefont {Saskin}, \citenamefont
  {Wilson}, \citenamefont {Alarcón}, \citenamefont {Greene},\ and\
  \citenamefont {Thompson}}]{burgers2021}%
  \BibitemOpen
  \bibfield  {author} {\bibinfo {author} {\bibfnamefont {A.~P.}\ \bibnamefont
  {Burgers}}, \bibinfo {author} {\bibfnamefont {S.}~\bibnamefont {Ma}},
  \bibinfo {author} {\bibfnamefont {S.}~\bibnamefont {Saskin}}, \bibinfo
  {author} {\bibfnamefont {J.}~\bibnamefont {Wilson}}, \bibinfo {author}
  {\bibfnamefont {M.~A.}\ \bibnamefont {Alarcón}}, \bibinfo {author}
  {\bibfnamefont {C.~H.}\ \bibnamefont {Greene}},\ and\ \bibinfo {author}
  {\bibfnamefont {J.~D.}\ \bibnamefont {Thompson}},\ }\href@noop {} {\bibinfo
  {title} {Controlling rydberg excitations using ion core transitions in
  alkaline earth atom tweezer arrays}} (\bibinfo {year} {2021}),\ \Eprint
  {https://arxiv.org/abs/2110.06902} {arXiv:2110.06902 [quant-ph]} \BibitemShut
  {NoStop}%
\bibitem [{\citenamefont {Pham}\ \emph {et~al.}(2021)\citenamefont {Pham},
  \citenamefont {Gallagher}, \citenamefont {Pillet}, \citenamefont {Lepoutre},\
  and\ \citenamefont {Cheinet}}]{pham2021}%
  \BibitemOpen
  \bibfield  {author} {\bibinfo {author} {\bibfnamefont {K.-L.}\ \bibnamefont
  {Pham}}, \bibinfo {author} {\bibfnamefont {T.}~\bibnamefont {Gallagher}},
  \bibinfo {author} {\bibfnamefont {P.}~\bibnamefont {Pillet}}, \bibinfo
  {author} {\bibfnamefont {S.}~\bibnamefont {Lepoutre}},\ and\ \bibinfo
  {author} {\bibfnamefont {P.}~\bibnamefont {Cheinet}},\ }\href@noop {}
  {\bibinfo {title} {A coherent light shift on alkaline-earth rydberg atoms
  from isolated core excitation without auto-ionization}} (\bibinfo {year}
  {2021}),\ \Eprint {https://arxiv.org/abs/2111.00982} {arXiv:2111.00982
  [physics.atom-ph]} \BibitemShut {NoStop}%
\bibitem [{\citenamefont {Norcia}\ \emph {et~al.}(2019)\citenamefont {Norcia},
  \citenamefont {Young}, \citenamefont {Eckner}, \citenamefont {Oelker},
  \citenamefont {Ye},\ and\ \citenamefont {Kaufman}}]{Norcia2019}%
  \BibitemOpen
  \bibfield  {author} {\bibinfo {author} {\bibfnamefont {M.~A.}\ \bibnamefont
  {Norcia}}, \bibinfo {author} {\bibfnamefont {A.~W.}\ \bibnamefont {Young}},
  \bibinfo {author} {\bibfnamefont {W.~J.}\ \bibnamefont {Eckner}}, \bibinfo
  {author} {\bibfnamefont {E.}~\bibnamefont {Oelker}}, \bibinfo {author}
  {\bibfnamefont {J.}~\bibnamefont {Ye}},\ and\ \bibinfo {author}
  {\bibfnamefont {A.~M.}\ \bibnamefont {Kaufman}},\ }\href
  {https://doi.org/10.1126/science.aay0644} {\bibfield  {journal} {\bibinfo
  {journal} {Science}\ }\textbf {\bibinfo {volume} {366}},\ \bibinfo {pages}
  {93} (\bibinfo {year} {2019})}\BibitemShut {NoStop}%
\bibitem [{\citenamefont {Madjarov}\ \emph {et~al.}(2019)\citenamefont
  {Madjarov}, \citenamefont {Cooper}, \citenamefont {Shaw}, \citenamefont
  {Covey}, \citenamefont {Schkolnik}, \citenamefont {Yoon}, \citenamefont
  {Williams},\ and\ \citenamefont {Endres}}]{madjarov2019}%
  \BibitemOpen
  \bibfield  {author} {\bibinfo {author} {\bibfnamefont {I.~S.}\ \bibnamefont
  {Madjarov}}, \bibinfo {author} {\bibfnamefont {A.}~\bibnamefont {Cooper}},
  \bibinfo {author} {\bibfnamefont {A.~L.}\ \bibnamefont {Shaw}}, \bibinfo
  {author} {\bibfnamefont {J.~P.}\ \bibnamefont {Covey}}, \bibinfo {author}
  {\bibfnamefont {V.}~\bibnamefont {Schkolnik}}, \bibinfo {author}
  {\bibfnamefont {T.~H.}\ \bibnamefont {Yoon}}, \bibinfo {author}
  {\bibfnamefont {J.~R.}\ \bibnamefont {Williams}},\ and\ \bibinfo {author}
  {\bibfnamefont {M.}~\bibnamefont {Endres}},\ }\href
  {https://doi.org/10.1103/PhysRevX.9.041052} {\bibfield  {journal} {\bibinfo
  {journal} {Phys. Rev. X}\ }\textbf {\bibinfo {volume} {9}},\ \bibinfo {pages}
  {041052} (\bibinfo {year} {2019})}\BibitemShut {NoStop}%
\bibitem [{\citenamefont {Schine}\ \emph {et~al.}(2021)\citenamefont {Schine},
  \citenamefont {Young}, \citenamefont {Eckner}, \citenamefont {Martin},\ and\
  \citenamefont {Kaufman}}]{schine2021}%
  \BibitemOpen
  \bibfield  {author} {\bibinfo {author} {\bibfnamefont {N.}~\bibnamefont
  {Schine}}, \bibinfo {author} {\bibfnamefont {A.~W.}\ \bibnamefont {Young}},
  \bibinfo {author} {\bibfnamefont {W.~J.}\ \bibnamefont {Eckner}}, \bibinfo
  {author} {\bibfnamefont {M.~J.}\ \bibnamefont {Martin}},\ and\ \bibinfo
  {author} {\bibfnamefont {A.~M.}\ \bibnamefont {Kaufman}},\ }\href@noop {}
  {\bibinfo {title} {Long-lived bell states in an array of optical clock
  qubits}} (\bibinfo {year} {2021}),\ \Eprint
  {https://arxiv.org/abs/2111.14653} {arXiv:2111.14653 [physics.atom-ph]}
  \BibitemShut {NoStop}%
\bibitem [{\citenamefont {Kuhr}\ \emph {et~al.}(2005)\citenamefont {Kuhr},
  \citenamefont {Alt}, \citenamefont {Schrader}, \citenamefont {Dotsenko},
  \citenamefont {Miroshnychenko}, \citenamefont {Rauschenbeutel},\ and\
  \citenamefont {Meschede}}]{kuhr2005}%
  \BibitemOpen
  \bibfield  {author} {\bibinfo {author} {\bibfnamefont {S.}~\bibnamefont
  {Kuhr}}, \bibinfo {author} {\bibfnamefont {W.}~\bibnamefont {Alt}}, \bibinfo
  {author} {\bibfnamefont {D.}~\bibnamefont {Schrader}}, \bibinfo {author}
  {\bibfnamefont {I.}~\bibnamefont {Dotsenko}}, \bibinfo {author}
  {\bibfnamefont {Y.}~\bibnamefont {Miroshnychenko}}, \bibinfo {author}
  {\bibfnamefont {A.}~\bibnamefont {Rauschenbeutel}},\ and\ \bibinfo {author}
  {\bibfnamefont {D.}~\bibnamefont {Meschede}},\ }\href
  {https://doi.org/10.1103/PhysRevA.72.023406} {\bibfield  {journal} {\bibinfo
  {journal} {Phys. Rev. A}\ }\textbf {\bibinfo {volume} {72}},\ \bibinfo
  {pages} {023406} (\bibinfo {year} {2005})}\BibitemShut {NoStop}%
\bibitem [{\citenamefont {D\"orscher}\ \emph {et~al.}(2018)\citenamefont
  {D\"orscher}, \citenamefont {Schwarz}, \citenamefont {Al-Masoudi},
  \citenamefont {Falke}, \citenamefont {Sterr},\ and\ \citenamefont
  {Lisdat}}]{dorscher2018}%
  \BibitemOpen
  \bibfield  {author} {\bibinfo {author} {\bibfnamefont {S.}~\bibnamefont
  {D\"orscher}}, \bibinfo {author} {\bibfnamefont {R.}~\bibnamefont {Schwarz}},
  \bibinfo {author} {\bibfnamefont {A.}~\bibnamefont {Al-Masoudi}}, \bibinfo
  {author} {\bibfnamefont {S.}~\bibnamefont {Falke}}, \bibinfo {author}
  {\bibfnamefont {U.}~\bibnamefont {Sterr}},\ and\ \bibinfo {author}
  {\bibfnamefont {C.}~\bibnamefont {Lisdat}},\ }\href
  {https://doi.org/10.1103/PhysRevA.97.063419} {\bibfield  {journal} {\bibinfo
  {journal} {Phys. Rev. A}\ }\textbf {\bibinfo {volume} {97}},\ \bibinfo
  {pages} {063419} (\bibinfo {year} {2018})}\BibitemShut {NoStop}%
\bibitem [{\citenamefont {Noguchi}\ \emph {et~al.}(2011)\citenamefont
  {Noguchi}, \citenamefont {Eto}, \citenamefont {Ueda},\ and\ \citenamefont
  {Kozuma}}]{noguchi2011}%
  \BibitemOpen
  \bibfield  {author} {\bibinfo {author} {\bibfnamefont {A.}~\bibnamefont
  {Noguchi}}, \bibinfo {author} {\bibfnamefont {Y.}~\bibnamefont {Eto}},
  \bibinfo {author} {\bibfnamefont {M.}~\bibnamefont {Ueda}},\ and\ \bibinfo
  {author} {\bibfnamefont {M.}~\bibnamefont {Kozuma}},\ }\href
  {https://doi.org/10.1103/PhysRevA.84.030301} {\bibfield  {journal} {\bibinfo
  {journal} {Phys. Rev. A}\ }\textbf {\bibinfo {volume} {84}},\ \bibinfo
  {pages} {030301} (\bibinfo {year} {2011})}\BibitemShut {NoStop}%
\bibitem [{\citenamefont {Beugnon}\ \emph {et~al.}(2007)\citenamefont
  {Beugnon}, \citenamefont {Tuchendler}, \citenamefont {Marion}, \citenamefont
  {Ga{\"e}tan}, \citenamefont {Miroshnychenko}, \citenamefont {Sortais},
  \citenamefont {Lance}, \citenamefont {Jones}, \citenamefont {Messin},
  \citenamefont {Browaeys},\ and\ \citenamefont {Grangier}}]{beugnon2007}%
  \BibitemOpen
  \bibfield  {author} {\bibinfo {author} {\bibfnamefont {J.}~\bibnamefont
  {Beugnon}}, \bibinfo {author} {\bibfnamefont {C.}~\bibnamefont {Tuchendler}},
  \bibinfo {author} {\bibfnamefont {H.}~\bibnamefont {Marion}}, \bibinfo
  {author} {\bibfnamefont {A.}~\bibnamefont {Ga{\"e}tan}}, \bibinfo {author}
  {\bibfnamefont {Y.}~\bibnamefont {Miroshnychenko}}, \bibinfo {author}
  {\bibfnamefont {Y.~R.~P.}\ \bibnamefont {Sortais}}, \bibinfo {author}
  {\bibfnamefont {A.~M.}\ \bibnamefont {Lance}}, \bibinfo {author}
  {\bibfnamefont {M.~P.~A.}\ \bibnamefont {Jones}}, \bibinfo {author}
  {\bibfnamefont {G.}~\bibnamefont {Messin}}, \bibinfo {author} {\bibfnamefont
  {A.}~\bibnamefont {Browaeys}},\ and\ \bibinfo {author} {\bibfnamefont
  {P.}~\bibnamefont {Grangier}},\ }\href {https://doi.org/10.1038/nphys698}
  {\bibfield  {journal} {\bibinfo  {journal} {Nature Physics}\ }\textbf
  {\bibinfo {volume} {3}},\ \bibinfo {pages} {696} (\bibinfo {year}
  {2007})}\BibitemShut {NoStop}%
\bibitem [{\citenamefont {Xia}\ \emph {et~al.}(2015)\citenamefont {Xia},
  \citenamefont {Lichtman}, \citenamefont {Maller}, \citenamefont {Carr},
  \citenamefont {Piotrowicz}, \citenamefont {Isenhower},\ and\ \citenamefont
  {Saffman}}]{xia2015}%
  \BibitemOpen
  \bibfield  {author} {\bibinfo {author} {\bibfnamefont {T.}~\bibnamefont
  {Xia}}, \bibinfo {author} {\bibfnamefont {M.}~\bibnamefont {Lichtman}},
  \bibinfo {author} {\bibfnamefont {K.}~\bibnamefont {Maller}}, \bibinfo
  {author} {\bibfnamefont {A.~W.}\ \bibnamefont {Carr}}, \bibinfo {author}
  {\bibfnamefont {M.~J.}\ \bibnamefont {Piotrowicz}}, \bibinfo {author}
  {\bibfnamefont {L.}~\bibnamefont {Isenhower}},\ and\ \bibinfo {author}
  {\bibfnamefont {M.}~\bibnamefont {Saffman}},\ }\href
  {https://doi.org/10.1103/PhysRevLett.114.100503} {\bibfield  {journal}
  {\bibinfo  {journal} {Phys. Rev. Lett.}\ }\textbf {\bibinfo {volume} {114}},\
  \bibinfo {pages} {100503} (\bibinfo {year} {2015})}\BibitemShut {NoStop}%
\bibitem [{\citenamefont {Guo}\ \emph {et~al.}(2020)\citenamefont {Guo},
  \citenamefont {He}, \citenamefont {Sheng}, \citenamefont {Yang},
  \citenamefont {Xu}, \citenamefont {Wang}, \citenamefont {Zhong},
  \citenamefont {Liu}, \citenamefont {Wang},\ and\ \citenamefont
  {Zhan}}]{guo2020}%
  \BibitemOpen
  \bibfield  {author} {\bibinfo {author} {\bibfnamefont {R.}~\bibnamefont
  {Guo}}, \bibinfo {author} {\bibfnamefont {X.}~\bibnamefont {He}}, \bibinfo
  {author} {\bibfnamefont {C.}~\bibnamefont {Sheng}}, \bibinfo {author}
  {\bibfnamefont {J.}~\bibnamefont {Yang}}, \bibinfo {author} {\bibfnamefont
  {P.}~\bibnamefont {Xu}}, \bibinfo {author} {\bibfnamefont {K.}~\bibnamefont
  {Wang}}, \bibinfo {author} {\bibfnamefont {J.}~\bibnamefont {Zhong}},
  \bibinfo {author} {\bibfnamefont {M.}~\bibnamefont {Liu}}, \bibinfo {author}
  {\bibfnamefont {J.}~\bibnamefont {Wang}},\ and\ \bibinfo {author}
  {\bibfnamefont {M.}~\bibnamefont {Zhan}},\ }\href
  {https://doi.org/10.1103/PhysRevLett.124.153201} {\bibfield  {journal}
  {\bibinfo  {journal} {Phys. Rev. Lett.}\ }\textbf {\bibinfo {volume} {124}},\
  \bibinfo {pages} {153201} (\bibinfo {year} {2020})}\BibitemShut {NoStop}%
\bibitem [{\citenamefont {Lemke}\ \emph {et~al.}(2009)\citenamefont {Lemke},
  \citenamefont {Ludlow}, \citenamefont {Barber}, \citenamefont {Fortier},
  \citenamefont {Diddams}, \citenamefont {Jiang}, \citenamefont {Jefferts},
  \citenamefont {Heavner}, \citenamefont {Parker},\ and\ \citenamefont
  {Oates}}]{lemke2009}%
  \BibitemOpen
  \bibfield  {author} {\bibinfo {author} {\bibfnamefont {N.~D.}\ \bibnamefont
  {Lemke}}, \bibinfo {author} {\bibfnamefont {A.~D.}\ \bibnamefont {Ludlow}},
  \bibinfo {author} {\bibfnamefont {Z.~W.}\ \bibnamefont {Barber}}, \bibinfo
  {author} {\bibfnamefont {T.~M.}\ \bibnamefont {Fortier}}, \bibinfo {author}
  {\bibfnamefont {S.~A.}\ \bibnamefont {Diddams}}, \bibinfo {author}
  {\bibfnamefont {Y.}~\bibnamefont {Jiang}}, \bibinfo {author} {\bibfnamefont
  {S.~R.}\ \bibnamefont {Jefferts}}, \bibinfo {author} {\bibfnamefont {T.~P.}\
  \bibnamefont {Heavner}}, \bibinfo {author} {\bibfnamefont {T.~E.}\
  \bibnamefont {Parker}},\ and\ \bibinfo {author} {\bibfnamefont {C.~W.}\
  \bibnamefont {Oates}},\ }\href
  {https://doi.org/10.1103/PhysRevLett.103.063001} {\bibfield  {journal}
  {\bibinfo  {journal} {Phys. Rev. Lett.}\ }\textbf {\bibinfo {volume} {103}},\
  \bibinfo {pages} {063001} (\bibinfo {year} {2009})}\BibitemShut {NoStop}%
\bibitem [{\citenamefont {Yamamoto}\ \emph {et~al.}(2016)\citenamefont
  {Yamamoto}, \citenamefont {Kobayashi}, \citenamefont {Kuno}, \citenamefont
  {Kato},\ and\ \citenamefont {Takahashi}}]{yamamoto2016}%
  \BibitemOpen
  \bibfield  {author} {\bibinfo {author} {\bibfnamefont {R.}~\bibnamefont
  {Yamamoto}}, \bibinfo {author} {\bibfnamefont {J.}~\bibnamefont {Kobayashi}},
  \bibinfo {author} {\bibfnamefont {T.}~\bibnamefont {Kuno}}, \bibinfo {author}
  {\bibfnamefont {K.}~\bibnamefont {Kato}},\ and\ \bibinfo {author}
  {\bibfnamefont {Y.}~\bibnamefont {Takahashi}},\ }\href
  {https://doi.org/10.1088/1367-2630/18/2/023016} {\bibfield  {journal}
  {\bibinfo  {journal} {New Journal of Physics}\ }\textbf {\bibinfo {volume}
  {18}},\ \bibinfo {pages} {023016} (\bibinfo {year} {2016})}\BibitemShut
  {NoStop}%
\bibitem [{\citenamefont {Porsev}\ \emph {et~al.}(1999)\citenamefont {Porsev},
  \citenamefont {Rakhlina},\ and\ \citenamefont {Kozlov}}]{porsev1999}%
  \BibitemOpen
  \bibfield  {author} {\bibinfo {author} {\bibfnamefont {S.~G.}\ \bibnamefont
  {Porsev}}, \bibinfo {author} {\bibfnamefont {Y.~G.}\ \bibnamefont
  {Rakhlina}},\ and\ \bibinfo {author} {\bibfnamefont {M.~G.}\ \bibnamefont
  {Kozlov}},\ }\href {https://doi.org/10.1103/PhysRevA.60.2781} {\bibfield
  {journal} {\bibinfo  {journal} {Phys. Rev. A}\ }\textbf {\bibinfo {volume}
  {60}},\ \bibinfo {pages} {2781} (\bibinfo {year} {1999})}\BibitemShut
  {NoStop}%
\bibitem [{\citenamefont {Barber}\ \emph {et~al.}(2008)\citenamefont {Barber},
  \citenamefont {Stalnaker}, \citenamefont {Lemke}, \citenamefont {Poli},
  \citenamefont {Oates}, \citenamefont {Fortier}, \citenamefont {Diddams},
  \citenamefont {Hollberg}, \citenamefont {Hoyt}, \citenamefont
  {Taichenachev},\ and\ \citenamefont {Yudin}}]{barber2008}%
  \BibitemOpen
  \bibfield  {author} {\bibinfo {author} {\bibfnamefont {Z.~W.}\ \bibnamefont
  {Barber}}, \bibinfo {author} {\bibfnamefont {J.~E.}\ \bibnamefont
  {Stalnaker}}, \bibinfo {author} {\bibfnamefont {N.~D.}\ \bibnamefont
  {Lemke}}, \bibinfo {author} {\bibfnamefont {N.}~\bibnamefont {Poli}},
  \bibinfo {author} {\bibfnamefont {C.~W.}\ \bibnamefont {Oates}}, \bibinfo
  {author} {\bibfnamefont {T.~M.}\ \bibnamefont {Fortier}}, \bibinfo {author}
  {\bibfnamefont {S.~A.}\ \bibnamefont {Diddams}}, \bibinfo {author}
  {\bibfnamefont {L.}~\bibnamefont {Hollberg}}, \bibinfo {author}
  {\bibfnamefont {C.~W.}\ \bibnamefont {Hoyt}}, \bibinfo {author}
  {\bibfnamefont {A.~V.}\ \bibnamefont {Taichenachev}},\ and\ \bibinfo {author}
  {\bibfnamefont {V.~I.}\ \bibnamefont {Yudin}},\ }\href
  {https://doi.org/10.1103/PhysRevLett.100.103002} {\bibfield  {journal}
  {\bibinfo  {journal} {Phys. Rev. Lett.}\ }\textbf {\bibinfo {volume} {100}},\
  \bibinfo {pages} {103002} (\bibinfo {year} {2008})}\BibitemShut {NoStop}%
\bibitem [{\citenamefont {Lochead}\ \emph {et~al.}(2013)\citenamefont
  {Lochead}, \citenamefont {Boddy}, \citenamefont {Sadler}, \citenamefont
  {Adams},\ and\ \citenamefont {Jones}}]{Lochead2013}%
  \BibitemOpen
  \bibfield  {author} {\bibinfo {author} {\bibfnamefont {G.}~\bibnamefont
  {Lochead}}, \bibinfo {author} {\bibfnamefont {D.}~\bibnamefont {Boddy}},
  \bibinfo {author} {\bibfnamefont {D.~P.}\ \bibnamefont {Sadler}}, \bibinfo
  {author} {\bibfnamefont {C.~S.}\ \bibnamefont {Adams}},\ and\ \bibinfo
  {author} {\bibfnamefont {M.~P.~A.}\ \bibnamefont {Jones}},\ }\href
  {https://doi.org/10.1103/PhysRevA.87.053409} {\bibfield  {journal} {\bibinfo
  {journal} {Phys. Rev. A}\ }\textbf {\bibinfo {volume} {87}},\ \bibinfo
  {pages} {053409} (\bibinfo {year} {2013})}\BibitemShut {NoStop}%
\bibitem [{\citenamefont {Nielsen}\ \emph {et~al.}(2020)\citenamefont
  {Nielsen}, \citenamefont {Rudinger}, \citenamefont {Proctor}, \citenamefont
  {Russo}, \citenamefont {Young},\ and\ \citenamefont
  {Blume-Kohout}}]{Nielsen2020}%
  \BibitemOpen
  \bibfield  {author} {\bibinfo {author} {\bibfnamefont {E.}~\bibnamefont
  {Nielsen}}, \bibinfo {author} {\bibfnamefont {K.}~\bibnamefont {Rudinger}},
  \bibinfo {author} {\bibfnamefont {T.}~\bibnamefont {Proctor}}, \bibinfo
  {author} {\bibfnamefont {A.}~\bibnamefont {Russo}}, \bibinfo {author}
  {\bibfnamefont {K.}~\bibnamefont {Young}},\ and\ \bibinfo {author}
  {\bibfnamefont {R.}~\bibnamefont {Blume-Kohout}},\ }\href@noop {} {\bibfield
  {journal} {\bibinfo  {journal} {Quantum Science and Technology}\ }\textbf
  {\bibinfo {volume} {5}},\ \bibinfo {pages} {044002} (\bibinfo {year}
  {2020})}\BibitemShut {NoStop}%
\bibitem [{\citenamefont {Lukin}\ \emph {et~al.}(2001)\citenamefont {Lukin},
  \citenamefont {Fleischhauer}, \citenamefont {Cote}, \citenamefont {Duan},
  \citenamefont {Jaksch}, \citenamefont {Cirac},\ and\ \citenamefont
  {Zoller}}]{lukin2001}%
  \BibitemOpen
  \bibfield  {author} {\bibinfo {author} {\bibfnamefont {M.~D.}\ \bibnamefont
  {Lukin}}, \bibinfo {author} {\bibfnamefont {M.}~\bibnamefont {Fleischhauer}},
  \bibinfo {author} {\bibfnamefont {R.}~\bibnamefont {Cote}}, \bibinfo {author}
  {\bibfnamefont {L.~M.}\ \bibnamefont {Duan}}, \bibinfo {author}
  {\bibfnamefont {D.}~\bibnamefont {Jaksch}}, \bibinfo {author} {\bibfnamefont
  {J.~I.}\ \bibnamefont {Cirac}},\ and\ \bibinfo {author} {\bibfnamefont
  {P.}~\bibnamefont {Zoller}},\ }\href
  {https://doi.org/10.1103/PhysRevLett.87.037901} {\bibfield  {journal}
  {\bibinfo  {journal} {Phys. Rev. Lett.}\ }\textbf {\bibinfo {volume} {87}},\
  \bibinfo {pages} {037901} (\bibinfo {year} {2001})}\BibitemShut {NoStop}%
\bibitem [{\citenamefont {Isenhower}\ \emph {et~al.}(2010)\citenamefont
  {Isenhower}, \citenamefont {Urban}, \citenamefont {Zhang}, \citenamefont
  {Gill}, \citenamefont {Henage}, \citenamefont {Johnson}, \citenamefont
  {Walker},\ and\ \citenamefont {Saffman}}]{isenhower2010}%
  \BibitemOpen
  \bibfield  {author} {\bibinfo {author} {\bibfnamefont {L.}~\bibnamefont
  {Isenhower}}, \bibinfo {author} {\bibfnamefont {E.}~\bibnamefont {Urban}},
  \bibinfo {author} {\bibfnamefont {X.~L.}\ \bibnamefont {Zhang}}, \bibinfo
  {author} {\bibfnamefont {A.~T.}\ \bibnamefont {Gill}}, \bibinfo {author}
  {\bibfnamefont {T.}~\bibnamefont {Henage}}, \bibinfo {author} {\bibfnamefont
  {T.~A.}\ \bibnamefont {Johnson}}, \bibinfo {author} {\bibfnamefont {T.~G.}\
  \bibnamefont {Walker}},\ and\ \bibinfo {author} {\bibfnamefont
  {M.}~\bibnamefont {Saffman}},\ }\href
  {https://doi.org/10.1103/PhysRevLett.104.010503} {\bibfield  {journal}
  {\bibinfo  {journal} {Phys. Rev. Lett.}\ }\textbf {\bibinfo {volume} {104}},\
  \bibinfo {pages} {010503} (\bibinfo {year} {2010})}\BibitemShut {NoStop}%
\bibitem [{\citenamefont {Jau}\ \emph {et~al.}(2016)\citenamefont {Jau},
  \citenamefont {Hankin}, \citenamefont {Keating}, \citenamefont {Deutsch},\
  and\ \citenamefont {Biedermann}}]{jau2016}%
  \BibitemOpen
  \bibfield  {author} {\bibinfo {author} {\bibfnamefont {Y.~Y.}\ \bibnamefont
  {Jau}}, \bibinfo {author} {\bibfnamefont {A.~M.}\ \bibnamefont {Hankin}},
  \bibinfo {author} {\bibfnamefont {T.}~\bibnamefont {Keating}}, \bibinfo
  {author} {\bibfnamefont {I.~H.}\ \bibnamefont {Deutsch}},\ and\ \bibinfo
  {author} {\bibfnamefont {G.~W.}\ \bibnamefont {Biedermann}},\ }\href
  {https://doi.org/10.1038/nphys3487} {\bibfield  {journal} {\bibinfo
  {journal} {Nature Physics}\ }\textbf {\bibinfo {volume} {12}},\ \bibinfo
  {pages} {71} (\bibinfo {year} {2016})}\BibitemShut {NoStop}%
\bibitem [{\citenamefont {Wilk}\ \emph {et~al.}(2010)\citenamefont {Wilk},
  \citenamefont {Ga\"etan}, \citenamefont {Evellin}, \citenamefont {Wolters},
  \citenamefont {Miroshnychenko}, \citenamefont {Grangier},\ and\ \citenamefont
  {Browaeys}}]{wilk2010}%
  \BibitemOpen
  \bibfield  {author} {\bibinfo {author} {\bibfnamefont {T.}~\bibnamefont
  {Wilk}}, \bibinfo {author} {\bibfnamefont {A.}~\bibnamefont {Ga\"etan}},
  \bibinfo {author} {\bibfnamefont {C.}~\bibnamefont {Evellin}}, \bibinfo
  {author} {\bibfnamefont {J.}~\bibnamefont {Wolters}}, \bibinfo {author}
  {\bibfnamefont {Y.}~\bibnamefont {Miroshnychenko}}, \bibinfo {author}
  {\bibfnamefont {P.}~\bibnamefont {Grangier}},\ and\ \bibinfo {author}
  {\bibfnamefont {A.}~\bibnamefont {Browaeys}},\ }\href
  {https://doi.org/10.1103/PhysRevLett.104.010502} {\bibfield  {journal}
  {\bibinfo  {journal} {Phys. Rev. Lett.}\ }\textbf {\bibinfo {volume} {104}},\
  \bibinfo {pages} {010502} (\bibinfo {year} {2010})}\BibitemShut {NoStop}%
\bibitem [{\citenamefont {Sackett}\ \emph {et~al.}(2000)\citenamefont
  {Sackett}, \citenamefont {Kielpinski}, \citenamefont {King}, \citenamefont
  {Langer}, \citenamefont {Meyer}, \citenamefont {Myatt}, \citenamefont {Rowe},
  \citenamefont {Turchette}, \citenamefont {Itano}, \citenamefont {Wineland},\
  and\ \citenamefont {Monroe}}]{sackett2000}%
  \BibitemOpen
  \bibfield  {author} {\bibinfo {author} {\bibfnamefont {C.~A.}\ \bibnamefont
  {Sackett}}, \bibinfo {author} {\bibfnamefont {D.}~\bibnamefont {Kielpinski}},
  \bibinfo {author} {\bibfnamefont {B.~E.}\ \bibnamefont {King}}, \bibinfo
  {author} {\bibfnamefont {C.}~\bibnamefont {Langer}}, \bibinfo {author}
  {\bibfnamefont {V.}~\bibnamefont {Meyer}}, \bibinfo {author} {\bibfnamefont
  {C.~J.}\ \bibnamefont {Myatt}}, \bibinfo {author} {\bibfnamefont
  {M.}~\bibnamefont {Rowe}}, \bibinfo {author} {\bibfnamefont {Q.~A.}\
  \bibnamefont {Turchette}}, \bibinfo {author} {\bibfnamefont {W.~M.}\
  \bibnamefont {Itano}}, \bibinfo {author} {\bibfnamefont {D.~J.}\ \bibnamefont
  {Wineland}},\ and\ \bibinfo {author} {\bibfnamefont {C.}~\bibnamefont
  {Monroe}},\ }\href {https://doi.org/10.1038/35005011} {\bibfield  {journal}
  {\bibinfo  {journal} {Nature}\ }\textbf {\bibinfo {volume} {404}},\ \bibinfo
  {pages} {256} (\bibinfo {year} {2000})}\BibitemShut {NoStop}%
\bibitem [{\citenamefont {de~L\'es\'eleuc}\ \emph {et~al.}(2018)\citenamefont
  {de~L\'es\'eleuc}, \citenamefont {Barredo}, \citenamefont {Lienhard},
  \citenamefont {Browaeys},\ and\ \citenamefont {Lahaye}}]{deleseleuc2018}%
  \BibitemOpen
  \bibfield  {author} {\bibinfo {author} {\bibfnamefont {S.}~\bibnamefont
  {de~L\'es\'eleuc}}, \bibinfo {author} {\bibfnamefont {D.}~\bibnamefont
  {Barredo}}, \bibinfo {author} {\bibfnamefont {V.}~\bibnamefont {Lienhard}},
  \bibinfo {author} {\bibfnamefont {A.}~\bibnamefont {Browaeys}},\ and\
  \bibinfo {author} {\bibfnamefont {T.}~\bibnamefont {Lahaye}},\ }\href
  {https://doi.org/10.1103/PhysRevA.97.053803} {\bibfield  {journal} {\bibinfo
  {journal} {Phys. Rev. A}\ }\textbf {\bibinfo {volume} {97}},\ \bibinfo
  {pages} {053803} (\bibinfo {year} {2018})}\BibitemShut {NoStop}%
\bibitem [{\citenamefont {Yang}\ \emph {et~al.}(2016)\citenamefont {Yang},
  \citenamefont {He}, \citenamefont {Guo}, \citenamefont {Xu}, \citenamefont
  {Wang}, \citenamefont {Sheng}, \citenamefont {Liu}, \citenamefont {Wang},
  \citenamefont {Derevianko},\ and\ \citenamefont {Zhan}}]{yang2016}%
  \BibitemOpen
  \bibfield  {author} {\bibinfo {author} {\bibfnamefont {J.}~\bibnamefont
  {Yang}}, \bibinfo {author} {\bibfnamefont {X.}~\bibnamefont {He}}, \bibinfo
  {author} {\bibfnamefont {R.}~\bibnamefont {Guo}}, \bibinfo {author}
  {\bibfnamefont {P.}~\bibnamefont {Xu}}, \bibinfo {author} {\bibfnamefont
  {K.}~\bibnamefont {Wang}}, \bibinfo {author} {\bibfnamefont {C.}~\bibnamefont
  {Sheng}}, \bibinfo {author} {\bibfnamefont {M.}~\bibnamefont {Liu}}, \bibinfo
  {author} {\bibfnamefont {J.}~\bibnamefont {Wang}}, \bibinfo {author}
  {\bibfnamefont {A.}~\bibnamefont {Derevianko}},\ and\ \bibinfo {author}
  {\bibfnamefont {M.}~\bibnamefont {Zhan}},\ }\href
  {https://doi.org/10.1103/PhysRevLett.117.123201} {\bibfield  {journal}
  {\bibinfo  {journal} {Phys. Rev. Lett.}\ }\textbf {\bibinfo {volume} {117}},\
  \bibinfo {pages} {123201} (\bibinfo {year} {2016})}\BibitemShut {NoStop}%
\bibitem [{\citenamefont {Levine}\ \emph {et~al.}(2021)\citenamefont {Levine},
  \citenamefont {Bluvstein}, \citenamefont {Keesling}, \citenamefont {Wang},
  \citenamefont {Ebadi}, \citenamefont {Semeghini}, \citenamefont {Omran},
  \citenamefont {Greiner}, \citenamefont {Vuletić},\ and\ \citenamefont
  {Lukin}}]{levine2021}%
  \BibitemOpen
  \bibfield  {author} {\bibinfo {author} {\bibfnamefont {H.}~\bibnamefont
  {Levine}}, \bibinfo {author} {\bibfnamefont {D.}~\bibnamefont {Bluvstein}},
  \bibinfo {author} {\bibfnamefont {A.}~\bibnamefont {Keesling}}, \bibinfo
  {author} {\bibfnamefont {T.~T.}\ \bibnamefont {Wang}}, \bibinfo {author}
  {\bibfnamefont {S.}~\bibnamefont {Ebadi}}, \bibinfo {author} {\bibfnamefont
  {G.}~\bibnamefont {Semeghini}}, \bibinfo {author} {\bibfnamefont
  {A.}~\bibnamefont {Omran}}, \bibinfo {author} {\bibfnamefont
  {M.}~\bibnamefont {Greiner}}, \bibinfo {author} {\bibfnamefont
  {V.}~\bibnamefont {Vuletić}},\ and\ \bibinfo {author} {\bibfnamefont
  {M.~D.}\ \bibnamefont {Lukin}},\ }\href@noop {} {\bibinfo {title} {Dispersive
  optical systems for scalable raman driving of hyperfine qubits}} (\bibinfo
  {year} {2021}),\ \Eprint {https://arxiv.org/abs/2110.14645} {arXiv:2110.14645
  [quant-ph]} \BibitemShut {NoStop}%
\bibitem [{\citenamefont {Levine}\ \emph {et~al.}(2018)\citenamefont {Levine},
  \citenamefont {Keesling}, \citenamefont {Omran}, \citenamefont {Bernien},
  \citenamefont {Schwartz}, \citenamefont {Zibrov}, \citenamefont {Endres},
  \citenamefont {Greiner}, \citenamefont {Vuleti\ifmmode~\acute{c}\else
  \'{c}\fi{}},\ and\ \citenamefont {Lukin}}]{levine2018}%
  \BibitemOpen
  \bibfield  {author} {\bibinfo {author} {\bibfnamefont {H.}~\bibnamefont
  {Levine}}, \bibinfo {author} {\bibfnamefont {A.}~\bibnamefont {Keesling}},
  \bibinfo {author} {\bibfnamefont {A.}~\bibnamefont {Omran}}, \bibinfo
  {author} {\bibfnamefont {H.}~\bibnamefont {Bernien}}, \bibinfo {author}
  {\bibfnamefont {S.}~\bibnamefont {Schwartz}}, \bibinfo {author}
  {\bibfnamefont {A.~S.}\ \bibnamefont {Zibrov}}, \bibinfo {author}
  {\bibfnamefont {M.}~\bibnamefont {Endres}}, \bibinfo {author} {\bibfnamefont
  {M.}~\bibnamefont {Greiner}}, \bibinfo {author} {\bibfnamefont
  {V.}~\bibnamefont {Vuleti\ifmmode~\acute{c}\else \'{c}\fi{}}},\ and\ \bibinfo
  {author} {\bibfnamefont {M.~D.}\ \bibnamefont {Lukin}},\ }\href
  {https://doi.org/10.1103/PhysRevLett.121.123603} {\bibfield  {journal}
  {\bibinfo  {journal} {Phys. Rev. Lett.}\ }\textbf {\bibinfo {volume} {121}},\
  \bibinfo {pages} {123603} (\bibinfo {year} {2018})}\BibitemShut {NoStop}%
\bibitem [{\citenamefont {Vaillant}\ \emph {et~al.}(2012)\citenamefont
  {Vaillant}, \citenamefont {Jones},\ and\ \citenamefont
  {Potvliege}}]{vaillant2012}%
  \BibitemOpen
  \bibfield  {author} {\bibinfo {author} {\bibfnamefont {C.~L.}\ \bibnamefont
  {Vaillant}}, \bibinfo {author} {\bibfnamefont {M.~P.~A.}\ \bibnamefont
  {Jones}},\ and\ \bibinfo {author} {\bibfnamefont {R.~M.}\ \bibnamefont
  {Potvliege}},\ }\href@noop {} {\bibfield  {journal} {\bibinfo  {journal}
  {Journal of Physics B: Atomic, Molecular and Optical Physics}\ }\textbf
  {\bibinfo {volume} {45}},\ \bibinfo {pages} {135004} (\bibinfo {year}
  {2012})}\BibitemShut {NoStop}%
\bibitem [{\citenamefont {Robicheaux}(2019)}]{robicheaux2019}%
  \BibitemOpen
  \bibfield  {author} {\bibinfo {author} {\bibfnamefont {F.}~\bibnamefont
  {Robicheaux}},\ }\href {https://doi.org/10.1088/1361-6455/ab4c22} {\bibfield
  {journal} {\bibinfo  {journal} {Journal of Physics B: Atomic, Molecular and
  Optical Physics}\ }\textbf {\bibinfo {volume} {52}},\ \bibinfo {pages}
  {244001} (\bibinfo {year} {2019})}\BibitemShut {NoStop}%
\bibitem [{\citenamefont {Covey}\ \emph {et~al.}(2019)\citenamefont {Covey},
  \citenamefont {Madjarov}, \citenamefont {Cooper},\ and\ \citenamefont
  {Endres}}]{covey2019}%
  \BibitemOpen
  \bibfield  {author} {\bibinfo {author} {\bibfnamefont {J.~P.}\ \bibnamefont
  {Covey}}, \bibinfo {author} {\bibfnamefont {I.~S.}\ \bibnamefont {Madjarov}},
  \bibinfo {author} {\bibfnamefont {A.}~\bibnamefont {Cooper}},\ and\ \bibinfo
  {author} {\bibfnamefont {M.}~\bibnamefont {Endres}},\ }\href
  {https://doi.org/10.1103/PhysRevLett.122.173201} {\bibfield  {journal}
  {\bibinfo  {journal} {Phys. Rev. Lett.}\ }\textbf {\bibinfo {volume} {122}},\
  \bibinfo {pages} {173201} (\bibinfo {year} {2019})}\BibitemShut {NoStop}%
\bibitem [{\citenamefont {Gil}\ \emph {et~al.}(2014)\citenamefont {Gil},
  \citenamefont {Mukherjee}, \citenamefont {Bridge}, \citenamefont {Jones},\
  and\ \citenamefont {Pohl}}]{gil2014}%
  \BibitemOpen
  \bibfield  {author} {\bibinfo {author} {\bibfnamefont {L.~I.~R.}\
  \bibnamefont {Gil}}, \bibinfo {author} {\bibfnamefont {R.}~\bibnamefont
  {Mukherjee}}, \bibinfo {author} {\bibfnamefont {E.~M.}\ \bibnamefont
  {Bridge}}, \bibinfo {author} {\bibfnamefont {M.~P.~A.}\ \bibnamefont
  {Jones}},\ and\ \bibinfo {author} {\bibfnamefont {T.}~\bibnamefont {Pohl}},\
  }\href {http://link.aps.org/doi/10.1103/PhysRevLett.112.103601} {\bibfield
  {journal} {\bibinfo  {journal} {Physical Review Letters}\ }\textbf {\bibinfo
  {volume} {112}},\ \bibinfo {pages} {103601} (\bibinfo {year}
  {2014})}\BibitemShut {NoStop}%
\bibitem [{\citenamefont {Kaubruegger}\ \emph {et~al.}(2019)\citenamefont
  {Kaubruegger}, \citenamefont {Silvi}, \citenamefont {Kokail}, \citenamefont
  {van Bijnen}, \citenamefont {Rey}, \citenamefont {Ye}, \citenamefont
  {Kaufman},\ and\ \citenamefont {Zoller}}]{kaubruegger2019}%
  \BibitemOpen
  \bibfield  {author} {\bibinfo {author} {\bibfnamefont {R.}~\bibnamefont
  {Kaubruegger}}, \bibinfo {author} {\bibfnamefont {P.}~\bibnamefont {Silvi}},
  \bibinfo {author} {\bibfnamefont {C.}~\bibnamefont {Kokail}}, \bibinfo
  {author} {\bibfnamefont {R.}~\bibnamefont {van Bijnen}}, \bibinfo {author}
  {\bibfnamefont {A.~M.}\ \bibnamefont {Rey}}, \bibinfo {author} {\bibfnamefont
  {J.}~\bibnamefont {Ye}}, \bibinfo {author} {\bibfnamefont {A.~M.}\
  \bibnamefont {Kaufman}},\ and\ \bibinfo {author} {\bibfnamefont
  {P.}~\bibnamefont {Zoller}},\ }\href
  {https://doi.org/10.1103/PhysRevLett.123.260505} {\bibfield  {journal}
  {\bibinfo  {journal} {Physical Review Letters}\ }\textbf {\bibinfo {volume}
  {123}},\ \bibinfo {pages} {260505} (\bibinfo {year} {2019})}\BibitemShut
  {NoStop}%
\bibitem [{\citenamefont {Jenkins}\ \emph {et~al.}()\citenamefont {Jenkins},
  \citenamefont {Lis}, \citenamefont {Senoo}, \citenamefont {McGrew},\ and\
  \citenamefont {Kaufman}}]{jenkins2021}%
  \BibitemOpen
  \bibfield  {author} {\bibinfo {author} {\bibfnamefont {A.}~\bibnamefont
  {Jenkins}}, \bibinfo {author} {\bibfnamefont {J.~W.}\ \bibnamefont {Lis}},
  \bibinfo {author} {\bibfnamefont {A.}~\bibnamefont {Senoo}}, \bibinfo
  {author} {\bibfnamefont {W.~F.}\ \bibnamefont {McGrew}},\ and\ \bibinfo
  {author} {\bibfnamefont {A.~M.}\ \bibnamefont {Kaufman}},\ }\href@noop {}
  {\bibinfo {title} {Ytterbium nuclear-spin qubits in an optical tweezer
  array}},\ \Eprint {https://arxiv.org/abs/Submitted (2021)} {Submitted (2021)}
  \BibitemShut {NoStop}%
\bibitem [{\citenamefont {Arora}\ \emph {et~al.}(2007)\citenamefont {Arora},
  \citenamefont {Safronova},\ and\ \citenamefont {Clark}}]{Arora2007}%
  \BibitemOpen
  \bibfield  {author} {\bibinfo {author} {\bibfnamefont {B.}~\bibnamefont
  {Arora}}, \bibinfo {author} {\bibfnamefont {M.~S.}\ \bibnamefont
  {Safronova}},\ and\ \bibinfo {author} {\bibfnamefont {C.~W.}\ \bibnamefont
  {Clark}},\ }\href {https://doi.org/10.1103/PhysRevA.76.052509} {\bibfield
  {journal} {\bibinfo  {journal} {Phys. Rev. A}\ }\textbf {\bibinfo {volume}
  {76}},\ \bibinfo {pages} {052509} (\bibinfo {year} {2007})}\BibitemShut
  {NoStop}%
\bibitem [{\citenamefont {Liao}\ \emph {et~al.}(1980)\citenamefont {Liao},
  \citenamefont {Freeman}, \citenamefont {Panock},\ and\ \citenamefont
  {Humphrey}}]{liao1980}%
  \BibitemOpen
  \bibfield  {author} {\bibinfo {author} {\bibfnamefont {P.}~\bibnamefont
  {Liao}}, \bibinfo {author} {\bibfnamefont {R.}~\bibnamefont {Freeman}},
  \bibinfo {author} {\bibfnamefont {R.}~\bibnamefont {Panock}},\ and\ \bibinfo
  {author} {\bibfnamefont {L.}~\bibnamefont {Humphrey}},\ }\href
  {https://doi.org/https://doi.org/10.1016/0030-4018(80)90013-9} {\bibfield
  {journal} {\bibinfo  {journal} {Optics Communications}\ }\textbf {\bibinfo
  {volume} {34}},\ \bibinfo {pages} {195} (\bibinfo {year} {1980})}\BibitemShut
  {NoStop}%
\bibitem [{\citenamefont {Beigang}\ \emph {et~al.}(1981)\citenamefont
  {Beigang}, \citenamefont {Matthias},\ and\ \citenamefont
  {Timmermann}}]{beigang1981}%
  \BibitemOpen
  \bibfield  {author} {\bibinfo {author} {\bibfnamefont {R.}~\bibnamefont
  {Beigang}}, \bibinfo {author} {\bibfnamefont {E.}~\bibnamefont {Matthias}},\
  and\ \bibinfo {author} {\bibfnamefont {A.}~\bibnamefont {Timmermann}},\
  }\href {https://doi.org/10.1103/PhysRevLett.47.326} {\bibfield  {journal}
  {\bibinfo  {journal} {Phys. Rev. Lett.}\ }\textbf {\bibinfo {volume} {47}},\
  \bibinfo {pages} {326} (\bibinfo {year} {1981})}\BibitemShut {NoStop}%
\bibitem [{\citenamefont {Aymar}(1984)}]{aymar1984}%
  \BibitemOpen
  \bibfield  {author} {\bibinfo {author} {\bibfnamefont {M.}~\bibnamefont
  {Aymar}},\ }\href
  {https://doi.org/https://doi.org/10.1016/0370-1573(84)90169-8} {\bibfield
  {journal} {\bibinfo  {journal} {Physics Reports}\ }\textbf {\bibinfo {volume}
  {110}},\ \bibinfo {pages} {163} (\bibinfo {year} {1984})}\BibitemShut
  {NoStop}%
\bibitem [{\citenamefont {Singer}\ \emph {et~al.}(2005)\citenamefont {Singer},
  \citenamefont {Stanojevic}, \citenamefont {Weidem{\"u}ller},\ and\
  \citenamefont {C{\^{o}}t{\'{e}}}}]{singer2005}%
  \BibitemOpen
  \bibfield  {author} {\bibinfo {author} {\bibfnamefont {K.}~\bibnamefont
  {Singer}}, \bibinfo {author} {\bibfnamefont {J.}~\bibnamefont {Stanojevic}},
  \bibinfo {author} {\bibfnamefont {M.}~\bibnamefont {Weidem{\"u}ller}},\ and\
  \bibinfo {author} {\bibfnamefont {R.}~\bibnamefont {C{\^{o}}t{\'{e}}}},\
  }\href {https://doi.org/10.1088/0953-4075/38/2/021} {\bibfield  {journal}
  {\bibinfo  {journal} {Journal of Physics B: Atomic, Molecular and Optical
  Physics}\ }\textbf {\bibinfo {volume} {38}},\ \bibinfo {pages} {S295}
  (\bibinfo {year} {2005})}\BibitemShut {NoStop}%
\bibitem [{\citenamefont {Saffman}\ \emph {et~al.}(2010)\citenamefont
  {Saffman}, \citenamefont {Walker},\ and\ \citenamefont
  {M\o{}lmer}}]{saffman2010}%
  \BibitemOpen
  \bibfield  {author} {\bibinfo {author} {\bibfnamefont {M.}~\bibnamefont
  {Saffman}}, \bibinfo {author} {\bibfnamefont {T.~G.}\ \bibnamefont
  {Walker}},\ and\ \bibinfo {author} {\bibfnamefont {K.}~\bibnamefont
  {M\o{}lmer}},\ }\href {https://doi.org/10.1103/RevModPhys.82.2313} {\bibfield
   {journal} {\bibinfo  {journal} {Rev. Mod. Phys.}\ }\textbf {\bibinfo
  {volume} {82}},\ \bibinfo {pages} {2313} (\bibinfo {year}
  {2010})}\BibitemShut {NoStop}%
\end{thebibliography}%
\bibliographystyle{apsrev4-2}
\end{document}